\documentclass{llncs}
\usepackage{mathtools}
\usepackage{makecell}
\usepackage{pdflscape}
\usepackage{rotating}
\usepackage{amsmath,amssymb}

\usepackage{textcomp,marvosym}

\usepackage{nameref,hyperref}

\usepackage[right]{lineno}

\usepackage{microtype}
\DisableLigatures[f]{encoding = *, family = * }

\usepackage{xargs}                      % Use more than one optional parameter in a new commands
\usepackage{array}

\usepackage{listings} 

\usepackage{subcaption}

\usepackage[british]{babel}
\usepackage{hhline}
\usepackage{multirow}

\usepackage{algorithmicx}
\usepackage{algpseudocode}
\usepackage{algorithm}
\algdef{SE}[DOWHILE]{Do}{doWhile}{\algorithmicdo}[1]{\algorithmicwhile\ #1}%

\usepackage{enumitem}
\let\llncssubparagraph\subparagraph
\let\subparagraph\paragraph
\usepackage[compact]{titlesec}
\let\subparagraph\llncssubparagraph

\usepackage{adjustbox}

\usepackage{xr}
\usepackage{catchfilebetweentags}

\usepackage{tkz-orm}
\usetikzlibrary{arrows,shapes,automata,backgrounds,petri,decorations.markings}
\usetikzlibrary{matrix,positioning,decorations.pathreplacing,calc,tikzmark}

\tikzset{
  ln/.style  = { draw, thick, fill=black, circle, inner sep=0.3mm},
  sqloc/.style  = { draw, thick, inner sep=1.0mm },
  loc/.style    = { draw, circle, thick, inner sep=1.0mm },
  notice/.style= { draw, rectangle callout, thick, rounded corners=5pt,fill=blue!10,callout relative pointer={#1} },
  treenode/.style = {align=center, inner sep=0pt, text centered,
    font=\sffamily},
  arn_n/.style = {treenode, rectangle, font=\sffamily\bfseries },% arbre rouge noir, noeud noir
  mymat/.style = { % matrix of math nodes,
    left delimiter={[}, right delimiter ={]},nodes={anchor=base east} },
  align at top/.style={baseline=(current bounding box.north)},
  stack/.style={rectangle split, rectangle split parts=#1,draw, anchor=center}
}

\usepackage{verbatim}
\usepackage{pgfplots}
\pgfplotsset{compat=1.15}
	
\def\cca#1{\cellcolor{blue!#10}\ifnum #1>5\color{white}\fi{#1}}

\usepackage{tikz}
\usetikzlibrary{tikzmark,decorations.pathreplacing,arrows,shapes,positioning,shadows,trees,shapes.gates.logic.US,arrows.meta,shapes,automata,petri,calc}
\usepackage{caption}
\usepackage{adjustbox}
\usepackage[colorinlistoftodos,prependcaption,textsize=tiny]{todonotes}
\newcommandx{\unsure}[2][1=]{\todo[linecolor=red,backgroundcolor=red!25,bordercolor=red,#1]{#2}}
\newcommandx{\change}[2][1=]{\todo[linecolor=blue,backgroundcolor=blue!25,bordercolor=blue,#1]{#2}}
\newcommandx{\info}[2][1=]{\todo[linecolor=OliveGreen,backgroundcolor=OliveGreen!25,bordercolor=OliveGreen,#1]{#2}}
\newcommandx{\improve}[2][1=]{\todo[linecolor=Plum,backgroundcolor=Plum!25,bordercolor=Plum,#1]{#2}}
\usepackage{verbatim}
\usepackage{scalefnt}
\usepackage{wrapfig}
\usepackage{pgfplots}
\usepackage{tikz-qtree}
\usepackage{pgfplots}

\mathchardef\mhyphen="2D
\newcommand{\shorteq}{%
  \settowidth{\@tempdima}{-}% Width of hyphen
  \resizebox{\@tempdima}{\height}{=}%
}

\colorlet{fv}{gray!55}
\colorlet{ai}{gray!10}
\colorlet{ar}{gray!38}

\usepackage[edges]{forest}
\usepackage[T1]{fontenc}

\tikzset{%
  parent/.style={align=center,text width=3cm,rounded corners=3pt},
  child/.style={align=center,text width=3cm,rounded corners=3pt},
}

\def\addlegendimage{\csname pgfplots@addlegendimage\endcsname}

 \usepackage{tikz}
 \usepackage{xparse}
\pgfdeclarelayer{myback}
\pgfsetlayers{myback,background,main}
\usetikzlibrary{tikzmark,decorations.pathreplacing,arrows,shapes,positioning,shadows,trees,shapes.gates.logic.US,arrows.meta,shapes,automata,petri,calc,matrix,backgrounds}
\tikzset{mycolor/.style = {line width=1bp,color=#1}}%
\tikzset{myfillcolor/.style = {draw,fill=#1}}%

\NewDocumentCommand{\highlight}{O{blue!40} m m}{%
\draw[mycolor=#1] (#2.north west)rectangle (#3.south east);
}

\NewDocumentCommand{\fhighlight}{O{blue!40} m m}{%
\draw[myfillcolor=#1] (#2.north west)rectangle (#3.south east);
}
 \usetikzlibrary{matrix,decorations.pathreplacing, calc, positioning}

\newcolumntype{+}{!{\vrule width 2pt}}

\newlength\savedwidth

\usepackage{array}
\newcolumntype{L}[1]{>{\raggedright\let\newline\\\arraybackslash\hspace{0pt}}m{#1}}
\newcolumntype{C}[1]{>{\centering\let\newline\\\arraybackslash\hspace{0pt}}m{#1}}
\newcolumntype{R}[1]{>{\raggedleft\let\newline\\\arraybackslash\hspace{0pt}}m{#1}}

\newcommand{\naturals}{\mathbb{N}}

\newcommand{\maps}{\rightarrow}
\newcommand{\pmaps}{\hookrightarrow}

\newcommand{\union}{{\cup} }

\newcommand{\powerset}[1]{2^{#1}}

\newcommand{\ltrue}{\mathbf{tt}}
\newcommand{\lfalse}{\mathbf{ff}}
\newcommand{\limplies}{\Rightarrow}

\newcommand{\Land}{\bigwedge}
\newcommand{\Lor}{\bigvee}

\newcommand{\lequiv}{\Leftrightarrow}
\newcommand{\landplus}{\mathrel{:\hspace{-3pt}\land\hspace{-3pt}=}}

\newcommand{\zthree}{\textsc{Z3}}
\newcommand{\ourtool}{\textsc{GlySynth}}

\newtheorem{df}{Definition}

\newcommand{\choosefinal}[2]{ #2 }  % arxiv version

\title{Automated inference of production rules for glycans}
\author{Ansuman Biswas$^2$, Ashutosh Gupta$^1$, Meghana Missula$^1$, and Mukund Thattai$^2$}
\institute{$^1$IITB and $^2$NCBS}

\date{May 2019}

\begin{document}

\maketitle

\begin{abstract}
  Glycans are tree-like polymers made up of sugar monomer building blocks. They are found on the surface of all living cells, and distinct glycan trees act as identity markers for distinct cell types. Proteins called GTase enzymes assemble glycans via the successive addition of monomer building blocks. The rules by which the enzymes operate are not fully understood. In this paper, we present the first SMT-solver-based iterative method that infers the assembly process of the glycans by analyzing the set of glycans from a cell. We have built a tool based on the method and applied it to infer rules based on published glycan data.
\end{abstract}

\section{Introduction}
\label{sec:intro}
The ability to control the assembly of small building blocks into large structures is of fundamental importance in biology and engineering. Familiar examples of this process from biology include the synthesis of linear DNA from nucleotide building blocks and the synthesis of linear proteins from amino-acid building blocks. In both these examples, the synthesis is templated: the new DNA or protein molecule is essentially copied from an existing molecule. However, most biological assembly proceeds without a template. For example, when an adult animal is grown from a fertilized egg, the genome within the egg contains a dynamical recipe encoding the animal rather than a template. The genome restricts and controls the set of events that can take place subsequent to fertilization.

While the process of animal development is too complex to study comprehensively, the same themes arise in the synthesis of complex tree-like sugar polymers known as glycans \cite{Varki2017} that are covalently attached to proteins. Unlike linear proteins and DNA, glycans are tree-like structures: their nodes are sugar monomers, and their edges are covalent carbon-carbon bonds. The tree-like structure of a glycan is a direct consequence of the fact that a sugar monomer can directly bond to at least three neighboring sugar monomers (in contrast to nucleotides or amino acids, which can only bind to two neighbors and are constrained to make a chain).

A given cell produces a specific set of glycan molecules that are present in the cell. Since different cells produce different sets of molecules,
the assembly process must be programmable: the assembly process includes a set of production rules. The reactions that underlie glycan production are carried out by enzymes known as GTases \cite{Varki2017}. There are hundreds of enzymes present in a given cell: each enzyme is a protein encoded by a distinct gene, which carries out a distinct biochemical reaction. The enzymes thus execute the production rules. A glycan tree is assembled piece by piece in successive steps. At each step, a production rule adds a small piece of a tree %(a monomer)
at the leaves or internal nodes of the current tree.
Not all the rules are applicable at all the leaves. The monomer at a leaf and current surroundings of the leaf controls
the applicability of a rule on the leaf.
Identifying the exact set of the production rules by extensive
biochemical experiments are costly and often needs an initial
hypothesis for the rules to test. Biologists must identify the production rules and their control
conditions by manually analyzing the set of observed glycans in a cell
and using prior knowledge of biochemistry.
We estimate that there may be more then $10^{70}$
possible rule sets if we consider all biological variations.
\footnote{For a problem having 10 monomers, 10 rules, 3 as rule size, 3 compartments
and fast-slow reactions, the search space is $\approx 10^{74}$ rules $(2^{10}*{(10+3-1) \choose (3-1)}*10^{(2^3 - 1)*10})$.}
This is an error-prone process since the production rules must generate
exactly the set of molecules in the cells and nothing else; and moreover, the data sets about which glycans are present in which cells are often incomplete. Manually comprehending all possible tree generation rules is difficult and ad-hoc. It would be useful to automate the process of learning which rules are operating in a given cell, based on incomplete data.

In this paper, we are presenting the {\em first} automated synthesis
method for the production rules.
Our method takes the observed glycan molecules in a cell as input and synthesizes
the possible production rules that may explain the observation.
To our knowledge,
our work is the {\em first} to consider the computational problem.
Our method of synthesis is similar to counterexample guided
inductive synthesis(CEGIS)~\cite{cegis}.
Several methods for solving problems of searching in a complex combinatorial
space use templates to define and limit their search space,
such as learning invariants of programs~\cite{InvGenTACAS09},
and synthesizing missing components in programs~\cite{sygus,Solar-Lezama2005}.
We also use templates to model the production rules.
% The templates usually have parameters that limit the search space.
% In our case, the size and number of templates are the parameters that
% limit sizes of pieces of trees that are added in one step,
% number of the production rules, and
% the depth of surrounding that are checked to control the rules.

% \begin{figure}[t]
%   \begin{tikzpicture}[thick,shorten >=1pt,node distance=2cm,on grid]
%     \node[sqloc]   (q0)                {Synthesis query};
%     \node   (qi) [xshift=-1cm, left=of q0]                {Input glycans};
%     \node[sqloc]           (q1) [xshift=2cm, right=of q0] {Counterexample query};
%     \node   (qo) [ below=of q1]                {Synthesized rules};
%     \node   (qf) [ below=of q0]                {Synthesis failed};
%     \path[->] (q0) edge [bend left=45] node [above] {Synthesized rules} node [below,xshift=-1.2cm,yshift=-3mm] {sat} (q1);
%     \path[->] (q1) edge [bend left=45] node [below] {Counterexample} node [above,xshift=1.2cm,yshift=3.5mm] {sat} (q0);
%     \path[->] (qi) edge  (q0);
%     \path[->] (q1) edge node [right,yshift=6mm] {unsat} (qo);
%     \path[->] (q0) edge node [left,yshift=6mm] {unsat} (qf);  
%   \end{tikzpicture}
%   \caption{The architecture of our synthesis method}
%   \label{fig:illus-method}
% \end{figure}

Our method is iterative.
%as illustrated in Figure~\ref{fig:illus-method}.
We first construct constraints encoding that a set of unknown rules
defined by templates can assemble the input set of molecules.
The generated constraints involve Boolean variables, finite range variables, and integer ordering
constraints.
We solve the constraints using an off-the-shelf SMT solver.
We call the query to the solver {\em synthesis query}.
If the constraints are unsatisfiable, there are no production rules with the search space
defined by the templates.
Otherwise, a solution of the constraints gives a set of rules.

However, there is also an additional requirement that a molecule
that is not in the input set must not be producible by the synthesized rules.
Therefore, the method looks for the producible molecules that are not in the input set.
Again the search of the molecule is assisted by a template, which bounds the height
of searched molecules.
We generate another set of constraints using the templates for the unknown molecule.
We again solve the constraints using an SMT solver.
We call the query to the solver {\em counterexample query}.
If there is no such molecule, our method reports
the synthesized rules.
Otherwise, we have found a producible molecule that is not in the input set.

We append our synthesis constraints with additional constraints stating that
no matter how we apply the synthesized rules, they will not produce the extra molecule.
Since there is a requirement that {\em all} possible applications of rules must satisfy
a condition, we have quantifiers in the constraints.
We use a solver that handles quantifiers over finite range variable in
the synthesis query.
We go to the next iteration of the method.
The method always terminates because the search space of rules is finite.
The set of rules synthesized need not be minimal or unique. The solver reports
the first set which satisfies the constraints.
However, our method is adoptable. We can add various optimization criteria to
find optimal rules for the given objectives, e.g., smallest rule sizes,
number of rules, etc.

Our encoding to constraints depends on the model of execution of the rules.
The current biological information is not sufficient to make a precise
and definite model, and do the synthesis.
We have also explored the variations of the models. For example, all rules
may apply simultaneously. They apply in batches because they
stay in different compartments.
The molecule under assembly may flow through the compartments.
The distribution of the stay of the molecules in a compartment also affects
the execution model.
Furthermore, we may have variations in the type and quality of data available to us. For example, we may have missed a produced molecule in experiments.
\choosefinal{We support the variations, which is presented in 
  the extended version of this paper~\cite{arxiv-this}.}{We support the variations.}

% \todo{Related work}
We have implemented the method in our tool~\ourtool. %The tool takes input data in its own format.
We have applied the tool on data sets from published sources (available in the database UniCarbKB \cite{Campbell2013}).
% We have obtained the rules for the data sets.
The output rules are within the expectations of biological
intuition.

% The key contributions of this paper are the following.
% \begin{itemize}
% \item We identified {\em a new area} of application for formal methods in biology.
% \item We have developed a {\em novel} synthesis method for discovering the production rules of glycan molecules from the output of the rules.
% \item We have implemented the method in a tool and applied it to data sets.
% \end{itemize}

We organize the paper as follows.
In Section~\ref{sec:bio}, we introduce the biological background.
%of the synthesis problem and present the basis of the formal model.
% If the reader wants to avoid getting into the biological details, she may skip the section.
In Section~\ref{sec:motivation}, we present a motivating example to illustrate our method.
In Section~\ref{sec:model}, we present the formal model of the glycans
and their production rules.
In Section~\ref{sec:algo}, we present our method for the synthesis problem.
%and the modifications in our methods to support the variations.
In Section~\ref{sec:experiments}-~\ref{sec:conclusion}, we present our experiments and conclude the paper. %on some glycan data.
% In Section~\ref{sec:conclusion}, we conclude the paper.
\choosefinal{}{We present the extended related work in Appendix~\ref{sec:related}.}

% We have added several appendix sections.
% In appendix~\ref{sec:variations}, we present the variations of the problem.
% In section~\ref{sec:related} and section~\ref{sec:conclusion},

%--------------------- DO NOT ERASE BELOW THIS LINE --------------------------

%%% Local Variables:
%%% mode: latex
%%% TeX-master: "main"
%%% End:

\section{Production of Glycans}
\label{sec:bio}

DNA and RNA are made by copying template DNA, in a process called transcription carried out by an enzyme called RNA polymerase. Proteins are made by copying a template messenger RNA, in a process called translation carried out by a molecular machine called a ribosome \cite{alberts2013essential}. In contrast, glycans are grown without a template, in a process called glycosylation. Glycosylation is carried out, not by a single enzyme, but by a large collection of so-called GTase enzymes that assemble one sugar monomer at a time into a final glycan tree. This process involves an ordered series of reactions, in which an enzyme first recruits the correct monomer, the enzyme-monomer complex binds to the target glycan at the appropriate motif, and finally a chemical reaction occurs which serves to bind the new monomer at the correct place on the glycan. The enzyme's binding motif can corresponding to a single monomer, or a large sub-structure of the entire glycan several nodes deep \cite{biswas2020promiscuity}. This process is reminiscent of a factory assembly line to make a car \cite{Jaiman2018}. However, the assembly process operates without a blueprint: the final glycan structure is determined by the behavior of the enzymes themselves.

The process of glycosylation is stochastic, governed by the Poisson statistics of single-step chemical reactions. One result of this stochasticity is that the enzymes can operate in different time orders \cite{Spahn2016}. It is as if factory workers could operate in many different orders while building the car, first adding doors and later windows. Moreover, the enzymes are promiscuous: they can add new monomers to many different places on the growing tree. This is as if the factory workers could add headlights at many different points on the car. Since there is no template, the existing tree determines where new monomers are added. Given the stochastic and promiscuous nature of the GTase enzymes, it is not surprising that the final product is highly variable \cite{Spahn2014}. The same set of enzymes can build many different glycan trees.

This variability is evident in the glycans observed to be produced by living cells. In a typical experiment, a protein is purified from a cell and the glycans attached to it are separated and their structure is characterized. Such an experiment produces a spectrum of glycan trees termed the protein's glycan profile \cite{Spahn2014}. A single glycan profile typically contains ten to twenty trees in measurable abundance, each tree being a tree of depth two to ten bonds. 

In \cite{Jaiman440792}, the authors had reported a method to infer the production rules when a single glycan is produced. However, the biologically interesting case is when the data set contains many glycan trees. This raises the following question: given a set of glycan trees produced by a cell,
can we infer the set of enzymes that produce the glycans? This is the problem we tackle here.

\begin{figure}[t]
  \centering
  \begin{minipage}{0.54\linewidth}
    \includegraphics[width=0.9\linewidth]{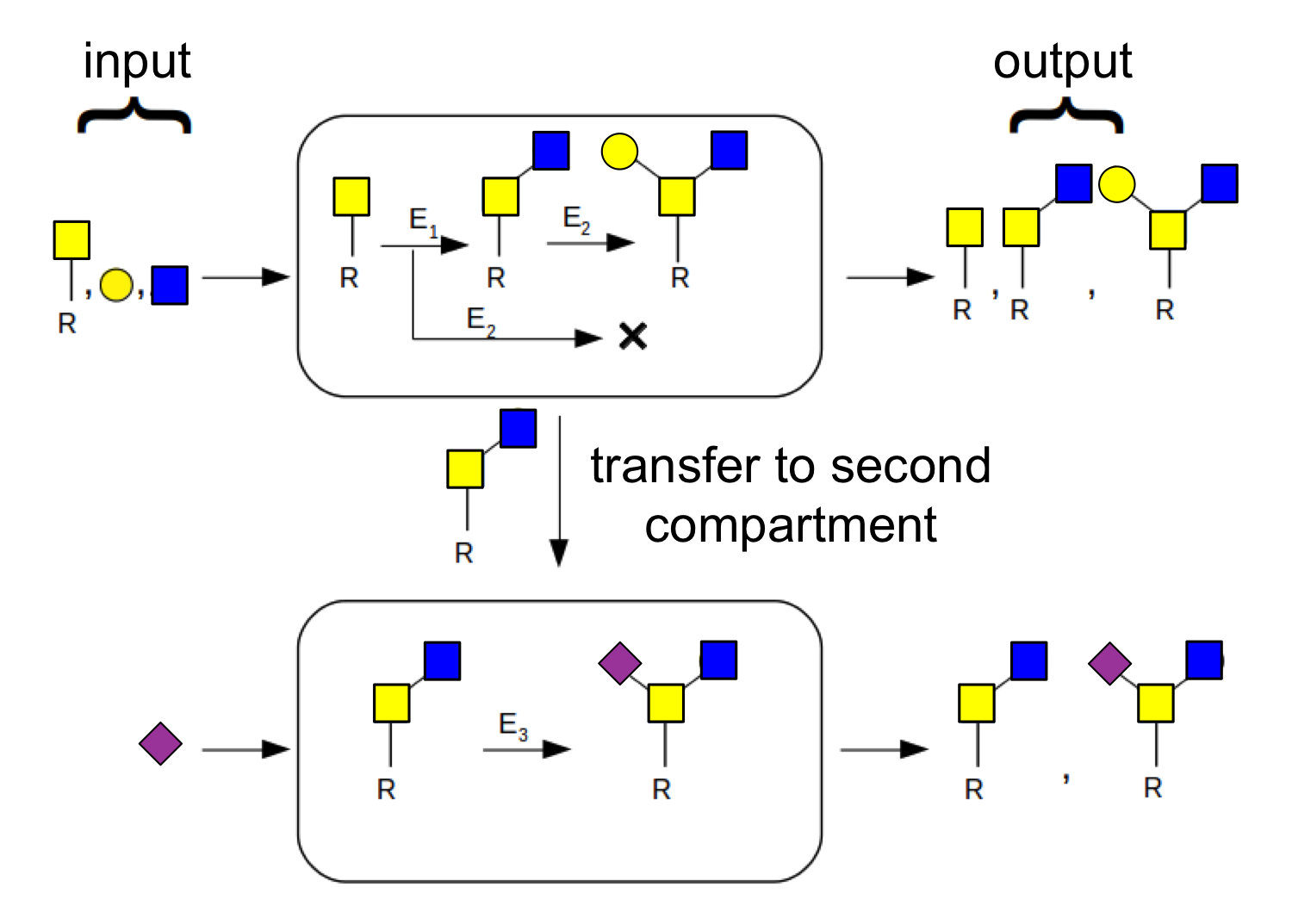}    
  \end{minipage}
  \begin{minipage}{0.44\linewidth}
    \caption{Biological details of glycan production. There are many types of sugar monomer building blocks; %(each represented by standard colored shapes);
      for example, GalNAc (yellow square), GlcNAc (blue square), Galactose (yellow circle), Sialic Acid (purple diamond), Fucose (red triangle) and so on \cite{Varki2017}.}
    \label{fig:glycan-rule}
  \end{minipage}
% \vspace{-9mm}
\end{figure}

In Figure~\ref{fig:glycan-rule}, we present details of glycan production. A glycan is a tree-like sugar tree (nodes linked by edges) attached to a substrate protein at the root (labeled `R'). Distinct edge orientations correspond to covalent bonds of distinct carbons on the sugar monomer. Curved boxes represent reaction compartments within cells, which are the site of glycan production. Each step of glycan growth (black arrows) represents the addition of a single new monomer to a specific attachment point on the tree. Each such step is catalyzed by an enzyme, labeled $E_i$. At any stage of growth, the tree can exit the reaction compartment as an output. Alternatively, it can be passed to a subsequent reaction compartment for further growth driven by different enzymes. Note that the enzymatic rule is sensitive to the two monomers being linked by a bond, as well as any branches. For example, enzyme $E_2$ will add a Galactose to a GalNAc only if the GlcNAc branch is present; otherwise, the reaction will not proceed (`X'). The structures, reactions, and enzymes shown here are illustrative, and they do not correspond to any measured data set; see the following section for a real example. In biological experiments, the combined outputs of every compartment are measured; the underlying reactions must be inferred.

%--------------------- DO NOT ERASE BELOW THIS LINE --------------------------

%%% Local Variables:
%%% mode: latex
%%% TeX-master: "main"
%%% End:

% \clearpage

\section{Motivating example}
\label{sec:motivation}
\begin{figure}[t]
\centering
\includegraphics[width=1\linewidth]{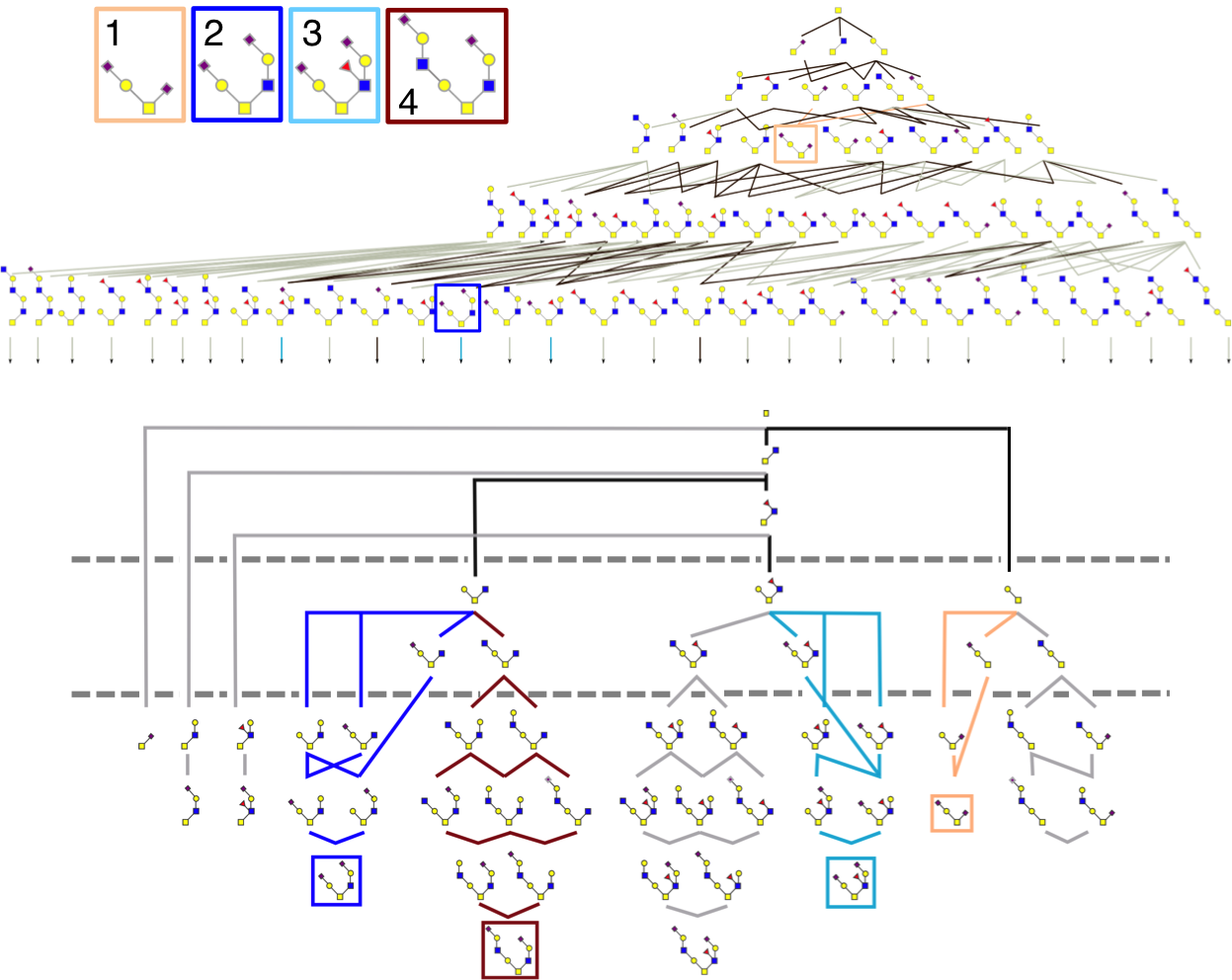}
\caption{A glycan data set. Figure credit: Anjali Jaiman, PhD thesis.}

\label{fig:dataset-gly}
\vspace{7.5mm}
% \vspace{-6mm}
\end{figure}

\begin{figure}[ht!]
% (sugar a 2)     #Box Yellow
% (sugar c 1)     #Circle yellow
% (sugar b 1)     #Box Blue
% (sugar d 0)     #Diamond red
  \small
  \begin{minipage}{0.5\linewidth}
    \mbox{}
    % \centering
  \begin{tikzpicture}[align at top, shorten >=1pt,thick,node distance=9mm,on grid]
    % (mol (a (c d) d) )
    \node[loc] (v0) {$A$};
    \node[loc, below left of=v0] (v1) {$C$};
    \node[loc, below of=v1] (v2) {$D$};
    \node[loc, below right of=v0] (v3) {$D$};

    \path[->] (v0) edge (v1);
    \path[->] (v1) edge (v2);
    \path[->] (v0) edge (v3);
  \end{tikzpicture}
  \qquad
  \begin{tikzpicture}[align at top, shorten >=1pt,thick,node distance=9mm,on grid]
    % (mol (a (c (b (c d)) ) (b (c d)) ) )
    \node[loc] (v0) {$A$};
    \node[loc, below left of=v0] (v1) {$C$};
    \node[loc, below of=v1] (v2) {$B$};
    \node[loc, below of=v2] (v6) {$C$};
    \node[loc, below of=v6] (v7) {$D$};
    \node[loc, below right of=v0] (v3) {$B$};
    \node[loc, below of=v3] (v4) {$C$};
    \node[loc, below of=v4] (v5) {$D$};

    \path[->] (v0) edge (v1);
    \path[->] (v1) edge (v2);
    \path[->] (v3) edge (v4);
    \path[->] (v4) edge (v5);
    \path[->] (v0) edge (v3);
    \path[->] (v2) edge (v6);
    \path[->] (v6) edge (v7);
  \end{tikzpicture}
  \vspace{-15mm}\par
  \begin{tikzpicture}[align at top, shorten >=1pt,thick,node distance=9mm,on grid]
    % (mol (a (c d) (b (c d)) ) )
    \node[loc] (v0) {$A$};
    \node[loc, below left of=v0] (v1) {$C$};
    \node[loc, below of=v1] (v2) {$D$};
    \node[loc, below right of=v0] (v3) {$B$};
    \node[loc, below of=v3] (v4) {$C$};
    \node[loc, below of=v4] (v5) {$D$};

    \path[->] (v0) edge (v1);
    \path[->] (v1) edge (v2);
    \path[->] (v0) edge (v3);
    \path[->] (v3) edge (v4);
    \path[->] (v4) edge (v5);
  \end{tikzpicture}
  \vspace{-3mm}
  \\
  \mbox{}\hfill(a)\hfill\mbox{}
\end{minipage}
\vrule
  \begin{minipage}{0.48\linewidth}
    \centering
    \begin{tikzpicture}[align at top, shorten >=1pt,thick,node distance=9mm,on grid]
      % 0:(a [(c d)] _)
      \node[loc] (v0) {$A$};
      \node[sqloc, below left of=v0] (v1) {$C$};
      \node[sqloc, below of=v1] (v2) {$D$};
      \path[->] (v0) edge (v1);
      \path[->] (v1) edge (v2);
    \end{tikzpicture}\qquad
    \begin{tikzpicture}[align at top, shorten >=1pt,thick,node distance=9mm,on grid]
      % 0:(a _ [(b-b c)])
      \node[loc] (v0) {$A$};
      \node[sqloc, below right of=v0] (v1) {$B$};
      \node[sqloc, below of=v1] (v2) {$C$};
      \path[->] (v0) edge (v1);
      \path[->] (v1) edge (v2);
    \end{tikzpicture}\qquad\hspace{2ex}
    \begin{tikzpicture}[align at top, shorten >=1pt,thick,node distance=9mm,on grid]
      % 0:(b-b (c [d]))
      \node[loc] (v0) {$B$};
      \node[loc, below of=v0] (v2) {$C$};
      \node[sqloc, below of=v2] (v3) {$D$};
      \path[->] (v0) edge (v2);
      \path[->] (v2) edge (v3);
    \end{tikzpicture}\\\vspace{2ex}
    \begin{tikzpicture}[align at top, shorten >=1pt,thick,node distance=9mm,on grid]
      % 0:(a _ [(d)])
      \node[loc] (v0) {$A$};
      \node[sqloc, below right of=v0] (v1) {$D$};
      \path[->] (v0) edge (v1);
    \end{tikzpicture}
    \quad
    \begin{tikzpicture}[align at top, shorten >=1pt,thick,node distance=9mm,on grid]
      % 0:(a [(c b-b)] (b-b _))
      \node[loc] (v0) {$A$};
      \node[loc, below right of=v0] (v1) {$B$};
      \node[sqloc, below left of=v0] (v2) {$C$};
      \node[sqloc, below of=v2] (v3) {$B$};
      \path[->] (v0) edge (v1);
      \path[->] (v0) edge (v2);
      \path[->] (v2) edge (v3);
    \end{tikzpicture}\qquad
    \begin{tikzpicture}[align at top, shorten >=1pt,thick,node distance=9mm,on grid]
      % 0:(c (b-b [c]))
      \node[loc] (v0) {$C$};
      \node[loc, below  of=v0] (v2) {$B$};
      \node[sqloc, below  of=v2] (v3) {$C$};
      \path[->] (v0) edge (v2);
      \path[->] (v2) edge (v3);
    \end{tikzpicture}
    \\
    (b)
  \end{minipage}
  \hrule
  \begin{minipage}{1.0\linewidth}
    \centering
  \begin{tikzpicture}[align at top, shorten >=1pt,thick,node distance=9mm,on grid]
    % (mol (a (c d) (b (c d)) ) )
    \node[loc] (v0) {$A$};
    % \node[loc, below right of=v0] (v3) {$B$};
    % \path[->] (v0) edge (v3);

    % \node[right of= v3, xshift=-0.2cm] (z1) {$\Rightarrow$};

    % \node[loc, right of= v0, xshift=2.1cm] (v0) {$A$};
    % \node[loc, below left of=v0] (v1) {$C$};
    % \node[loc, below right of=v0] (v3) {$B$};
    % \path[->] (v0) edge (v1);
    % \path[->] (v0) edge (v3);

    \node[right of= v0, yshift=-0.7cm, xshift=-0.2cm] (z2) {$\Rightarrow$};

    \node[loc, right of= v0, xshift=2.1cm] (v0) {$A$};
    \node[loc, below left of=v0] (v1) {$C$};
    \node[loc, below of=v1] (v2) {$D$};
    % \node[loc, below right of=v0] (v3) {$B$};
    \path[->] (v0) edge (v1);
    \path[->] (v1) edge (v2);
    % \path[->] (v0) edge (v3);

    \node[right of= v0, yshift=-0.7cm, xshift=-0.2cm] (z3) {$\Rightarrow$};

    \node[loc, right of= v0, xshift=2.1cm] (v0) {$A$};
    \node[loc, below left of=v0] (v1) {$C$};
    \node[loc, below of=v1] (v2) {$D$};
    \node[loc, below right of=v0] (v3) {$B$};
    \node[loc, below of=v3] (v4) {$C$};
    \path[->] (v0) edge (v1);
    \path[->] (v1) edge (v2);
    \path[->] (v0) edge (v3);
    \path[->] (v3) edge (v4);

    \node[right of= v3, xshift=-0.2cm] (z4) {$\Rightarrow$};

    \node[loc, right of= v0, xshift=2.1cm] (v0) {$A$};
    \node[loc, below left of=v0] (v1) {$C$};
    \node[loc, below of=v1] (v2) {$D$};
    \node[loc, below right of=v0] (v3) {$B$};
    \node[loc, below of=v3] (v4) {$C$};
    \node[loc, below of=v4] (v5) {$D$};
    \path[->] (v0) edge (v1);
    \path[->] (v1) edge (v2);
    \path[->] (v0) edge (v3);
    \path[->] (v3) edge (v4);
    \path[->] (v4) edge (v5);

    % % \node[ below of = z2, yshift=-1cm] (b1) {};
    % \draw [decorate,decoration={brace,amplitude=7pt}]
    % (-0.5,0.5) -- ++(4.5cm,0) node [midway,above,yshift=3pt] {Compartment 1};
    % \draw [decorate,decoration={brace,amplitude=7pt}]
    % (5,0.5) -- ++(8.5cm,0) node [midway,above,yshift=3pt] {Compartment 2};
  \end{tikzpicture}\\
  \vspace{-3mm}
  (c)
\end{minipage}
\hrule
\begin{minipage}{0.2\linewidth}
  \centering
  \begin{tikzpicture}[align at top, shorten >=1pt,thick,node distance=1cm,on grid]
    \node[loc] (v0) {$A$};
    \node[loc, below left of=v0] (v1) {$C$};
    \node[loc, below of=v1] (v2) {$B$};
    \node[loc, below right of=v0] (v3) {$D$};
    \path[->] (v0) edge (v1);
    \path[->] (v1) edge (v2);
    \path[->] (v0) edge (v3);
  \end{tikzpicture}\\
  % \vspace{12ex}
  (d)
\end{minipage}
\vrule
\begin{minipage}{0.75\linewidth}
  \centering
  \begin{tikzpicture}[align at top, shorten >=1pt,thick,node distance=1cm,on grid]
    % 0:(a [(c _)] _)    
    \node[loc] (v0) {$A$};
    \node[sqloc, below left of=v0] (v1) {$C$};
    \path[->] (v0) edge (v1);
  \end{tikzpicture}\qquad
  \begin{tikzpicture}[align at top, shorten >=1pt,thick,node distance=1cm,on grid]
    % 0:(a _ [(b-b c)])
    \node[loc] (v0) {$A$};
    \node[sqloc, below right of=v0] (v1) {$B$};
    \node[sqloc, below of=v1] (v2) {$C$};
    \path[->] (v0) edge (v1);
    \path[->] (v1) edge (v2);
  \end{tikzpicture}\qquad
  \begin{tikzpicture}[align at top, shorten >=1pt,thick,node distance=1cm,on grid]
    % 0:(a _ [(d)])
    \node[loc] (v0) {$A$};
    \node[sqloc, below right of=v0] (v1) {$D$};
    \path[->] (v0) edge (v1);
  \end{tikzpicture}
  % \vspace{-6mm}
  % \\
  \quad
  \begin{tikzpicture}[align at top, shorten >=1pt,thick,node distance=1cm,on grid]
    % 0:(c [(b-b c)])
    \node[loc] (v0) {$C$};
    \node[sqloc, below of=v0] (v1) {$B$};
    \node[sqloc, below of=v1] (v2) {$C$};
    \path[->] (v0) edge (v1);
    \path[->] (v1) edge (v2);
  \end{tikzpicture}\quad
  \begin{tikzpicture}[align at top, shorten >=1pt,thick,node distance=1cm,on grid]
    % 0:(c [(d)])
    \node[loc] (v0) {$C$};
    \node[sqloc, below of=v0] (v1) {$D$};
    \path[->] (v0) edge (v1);
  \end{tikzpicture}\quad
  \begin{tikzpicture}[align at top, shorten >=1pt,thick,node distance=1cm,on grid]
    % 0:[(a (b-b _) _)]
    \node[loc] (v0) {$A$};
    \node[sqloc, below left of=v0] (v1) {$B$};
    \path[->] (v0) edge (v1);
  \end{tikzpicture}
  % % 0:(a _ [b])
  % % 0:(a _ [d])
  % % 1:(a [c] _)
  % % 1:(c [b])
  % % 1:(c [d])
  % % 1:(b [c])
  %   \centering
  %   Rules in compartment 1\\
  %   \begin{tikzpicture}[align at top, shorten >=1pt,thick,node distance=1cm,on grid]
  %     % 0:(a _ [b])
  %     \node[loc] (v0) {$A$};
  %     \node[sqloc, below right of=v0] (v3) {$B$};
  %     \path[->] (v0) edge (v3);
  %   \end{tikzpicture}\quad
  %   \begin{tikzpicture}[align at top, shorten >=1pt,thick,node distance=1cm,on grid]
  %     % 0:(a _ [d])
  %     \node[loc] (v0) {$A$};
  %     \node[sqloc, below right of=v0] (v1) {$D$};
  %     \path[->] (v0) edge (v1);
  %   \end{tikzpicture}\\\vspace{2ex}
  %   Rules in compartment 2\\
  %   \begin{tikzpicture}[align at top, shorten >=1pt,thick,node distance=1cm,on grid]
  %     % 0:(c [b])
  %     \node[loc] (v0) {$C$};
  %     \node[sqloc, below of=v0] (v1) {$B$};
  %     \path[->] (v0) edge (v1);
  %   \end{tikzpicture}\qquad
  %   \begin{tikzpicture}[align at top, shorten >=1pt,thick,node distance=1cm,on grid]
  %     % 1:(a [c] _)
  %     \node[loc] (v0) {$A$};
  %     \node[sqloc, below left of=v0] (v1) {$C$};
  %     \path[->] (v0) edge (v1);
  %   \end{tikzpicture}\qquad
  %   \begin{tikzpicture}[align at top, shorten >=1pt,thick,node distance=1cm,on grid]
  %     % 1:(c [d])
  %     \node[loc] (v0) {$C$};
  %     \node[sqloc, below of=v0] (v1) {$D$};
  %     \path[->] (v0) edge (v1);
  %   \end{tikzpicture}\qquad
  %   \begin{tikzpicture}[align at top,shorten >=1pt,thick,node distance=1cm,on grid]
  %     % 1:(b [c])
  %     \node[loc] (v0) {$B$};
  %     \node[sqloc, below of=v0] (v1) {$C$};
  %     \path[->] (v0) edge (v1);
  %   \end{tikzpicture}\\\vspace{2ex}
    (e)
  \end{minipage}

% \vrule
% \begin{minipage}{0.3\linewidth}
%   \centering
%   \begin{tikzpicture}[align at top, shorten >=1pt,thick,node distance=1cm,on grid]
%     \node[loc] (v0) {$A$};
%     \node[loc, below left of=v0] (v1) {$C$};
%     \node[loc, below of=v1] (v2) {$B$};
%     \node[loc, below of=v2] (v6) {$C$};
%     \node[loc, below of=v6] (v7) {$B$};
%     % \node[loc, below right of=v0] (v3) {$B$};

%     \path[->] (v0) edge (v1);
%     \path[->] (v1) edge (v2);
%     % \path[->] (v3) edge (v4);
%     % \path[->] (v4) edge (v5);
%     % \path[->] (v0) edge (v3);
%     \path[->] (v2) edge (v6);
%     \path[->] (v6) edge (v7);
%     \draw [decorate,decoration={brace,amplitude=7pt}]
%     (-0.3cm,-2cm) -- ++(0cm,-2cm) node [midway,right,xshift=2mm]
%     {Repeated pattern};
%   \end{tikzpicture}\\
%   (e)
% \end{minipage}
  
  \vspace{-3mm}
  \caption{(a) A schematic example of a data set that includes three glycan oligomers.
    % Note that we have used a convention of keeping the root on top, which is opposite to the biological convention of keeping the root at the bottom.
    (b) A set of production rules for the glycan molecules
    (c) The steps of producing the middle glycan molecule
    (d) An undesired molecule.
    % The conditions on siblings help us avoid producing it.
    % (e) Another undesired molecule.
    % The conditions on ancestors help us avoid producing it.
    (e) The synthesized rules at the first iteration.
  }
  
  \label{fig:ex-gly}
  \vspace{6mm}
  % \vspace{-9mm}
\end{figure}

%%% Local Variables:
%%% mode: latex
%%% TeX-master: "main"
%%% End:

In this section, we first present a motivating example to illustrate our method.
In Figure~\ref{fig:dataset-gly},
we consider the glycan oligomers associated with human chorionic
gonadotropin~\cite{Harrd1992}.
The data set has four glycan oligomers (shown in boxes and numbered). We assume that all these oligomers
are built by starting from a root GalNAc (yellow square) by adding one monomer at a
time (lines between glycans represent monomer addition reactions). At the top of the figure, we illustrate
if all enzymes (rules) operate in a
single compartment, a large number of glycans can potentially be made in addition to the measured ones.
In the lower part of the figure, we illustrate if the enzymes are split into three compartments
(separated by dotted lines), then certain reactions are prevented from occurring. Thus, reducing the set of structures.
In this case, we assume that only the terminal (bottom-most) structures will be produced as outputs.
Here we have assumed certain rules of operation that are most consistent with the observed glycan data set.
The goal of this paper is to infer the rules.
% \vspace{-2.5mm}

Now we consider the synthesis problem. In Figure~\ref{fig:ex-gly}(a), we present a set of
glycan molecules present in a cell consists of three molecules, which are structurally similar to the three glycan molecules in Figure~\ref{fig:dataset-gly}.
To keep illustration simple, we have dropped the
third glycan molecule from the earlier set.
The molecules contain four types of monomers.
As we are considering the abstract case, we have named them $A$, $B$, $C$, $D$.
Each monomer is associated with an arity, i.e.,
the maximum number of potential children.
The arities of the monomers are $2$, $1$, $1$, and $0$, respectively.

% Before considering the synthesis problem that produces exactly
% the three molecules,

Let us first consider six rules in Figure~\ref{fig:ex-gly}(b)
that produce the molecules.
All the rules are in the same compartment, i.e., they can be applied in arbitrary order.
The rules have two kinds of nodes.
If the circular nodes are present around a node,
the rule is enabled and may append the molecule at the
node with the square nodes.
In Figure~\ref{fig:ex-gly}(c), we show the steps of generating the last
glycan molecule.
The first two steps add two nodes at a time.
The last step looks at the two ancestors before adding a single node.
% We may avoid the need for compartments if the operations can add multiple nodes at
% a time or look for a deep context for being enabled.

The second last rule in Figure~\ref{fig:ex-gly}(b) has a non-trivial condition
on the sibling of the anchor leaf node.
It requires, the parent of the new node should be $A$ and the right sibling
must be $B$.
If we do not have the sibling condition, we may be able to construct the molecule
in Figure~\ref{fig:ex-gly}(d) using the fourth and the modified fifth rule.
The molecule is not in a subtree of any of the three input glycan molecules.
Therefore, there are scenarios where rules must look into the context before applying
themselves.

% Furthermore, there is another interesting case occurs in this example.
% Please look at the last two rules.
% The second last rule adds $C\rightarrow B$ branch.
% The last rule checks if $C\rightarrow B$ nodes are present then adds $C$.
% If both rules were adding one node at a time without checking deep context,
% then the undesired molecule pattern in Figure~\ref{fig:ex-gly}(e) is
% also producible.
% We would have needed compartments to avoid the undesirable molecule.
% Therefore, there is a tread-off between rule sizes and the number of compartments.

Our method for synthesis takes the three glycan molecules as input.
It also needs the budget of resources to search for the production rules.
If we allow an arbitrary number of rules,
and the rules to look at their context up to
an arbitrary depth, % and an arbitrarily large number of compartments,
then there is a trivial solution.
Therefore, our method limits the number and size of rules.
% parameters.
% The first parameter limits the number of rules.
% The second parameter limits the maximum height of the rule trees.
% The third parameter limits the maximum number of compartments.
% If we give larger values to the parameters, then we have more chances to
% find the production rules.
For this illustration, we searched for the seven production rules with
three as the limit on the rule heights.
All rules are in a single compartment.

\ourtool, the tool that implements the method, reported the synthesized rules from
Figure~\ref{fig:ex-gly}(b) in 0.85 seconds.
In our tool, we first construct a synthesis query using the templates to encode
that a set of rules produces the input molecules.
We call a solver to solve the synthesis query.
After the first query, we obtain the rules presented in Figure~\ref{fig:ex-gly}(e).
The rule set can produce molecules that are not in input.
% Next, we look for another molecule that may be produced by the rule.
% Using another query to the solver, we find a counterexample molecule that is
% producible by the rules but is not in the input set.
% Coincidently, we have already presented the counterexample molecule in Figure~\ref{fig:ex-gly}(e).
% Afterword, we add constraints that the molecule must not
% be producible by the synthesized rules in our synthesis query and
% go for the next iteration.
We need to iterate further.
After 8 iterations, our tool synthesizes a set of rules that
satisfies the requirements.

%--------------------- DO NOT ERASE BELOW THIS LINE --------------------------

%%% Local Variables:
%%% mode: latex
%%% TeX-master: "main"
%%% End:

\section{Modelling of the synthesis problem}
\label{sec:model}

In this section, we present the formal model for the synthesis problem.
We model glycan molecules and production rules as labeled trees. The glycan molecules are assembled by applying the production rules repeatedly.
Our synthesis problem reduces into finding the pieces of trees that represent the production rules.

Let $S$ be the set of sugar monomers that builds glycans,
the oligomer molecules. Each $s \in S$ is associated with arity $m$ 
(written $arity(s) = m$).
% , {\em i.e.}, the maximum number of children for $s$.
The children of the monomers are indexed. We refer to the $k$th child of $s$ for some $k \leq arity(s)$.
They correspond to bonds at specific positions in the monomers where children are connected.
Now we define the glycan molecules as labeled trees. Now onward we refer to the glycans simply as molecules.

\begin{df}
A {\em molecule} $m = (V,M,C,v_0)$ is a labeled tree, where 
$V$ is a set of nodes in the tree,
$M : V \maps S$ maps nodes to their label, 
$C : V \times \naturals \pmaps V$ maps the indexed children of nodes, and
$v_0 \in V$ is the root of $m$.
A molecule must respect the arity of monomers, i.e., if $M(v) = s$ and $C(v,n) = v'$ then $n \leq arity(s)$.
\end{df}
% \todo{Rewrite the following paragraph}
% \todo{definitions mixed with algorithms seems strange I think you should define all the formal stuff first and then present the algorithms}
Let us define notations related to the tree structure.
Let $\mathit{m = (V,M,C,v_0)}$ and $m' = (V',M',C',v_0')$ be molecules.
With an abuse of notation, we write $v \in m$ to denote $v \in V$.
For each $v \in m$, if $(v,n)$ is not in the domain of $C$, we write $C(v,n) = \bot$.
We assume that $C(v,0) = \bot$.
%and if $n > n' > 0$ and $C(v,n) \neq \bot$, then $C(v,n') \neq \bot$.
Let $NumberOfChildren(v)$ be equal to the number of $n$s such that $C(v,n) \neq \bot$. 
A node $v \in V$ is a {\em leaf} of $m$ if $C(v,n) = \bot$ for each $n$.
Let $depth(v)$ be the length of the path from $v_0$ to $v$.
A {\em branch} of $m$ is a path from $v_0$ to some leaf of $m$.
Let $height(m)$ be the length of the longest branch in $m$.
We define ancestor relation recursively as follows. Let $ancestor(m,v,0) = v$.
For for $d > 0$, let $ancestor(m,v,d) = ancestor(m,v', d-1)$ if $C(v',i) = v$ for some $i$.

Since we will be matching the parts of the trees and applying rules to expand them,
let us introduce notations for matching.
Let recursively-defined predicate $Match(m,v,m',v')$ state that 
$v \in m$, $v' \in m'$, $s = M(v) = M(v')$, and $Match( m, C(v,n), m', C(v',n) )$ for each $n \leq arity(s)$ such that $C(v,n) \neq \bot$.
In other words, the subtree in $m$ rooted at $v$ is embedded in $m'$ at node $v'$.
Let $subtree(m)$ be the set of molecules such that
$m' = (\_,\_,\_,v_0') \in subtree(m) \lequiv Match(m,v_0,m',v_0')$.
Let us also define a utility to copy a subtree of a molecule into another molecule.
Let $Copy(m,v)$ return a molecule $m'' = (V'',M'',C'',v_0'')$ such that 
$V''$ is a set of fresh nodes, $Match(m,v,m'',v_0'')$, and $Match(m'',v_0'',m,v)$.
Both the $Match$ conditions say that the trees rooted at
$v$ and $v_0''$ are identical.

Now we define the model of the production process of the molecules.
A production rule expands a molecule $m$ by attaching a new piece of tree at a node that has a vacant spot among its children if the surroundings of the
node satisfy some condition.
The rule is modeled as a tree that has two parts.
One part should already be there in $m$ and the other part will be appended to $m$.

\begin{df}
  A {\em production rule} $r = (V, M, C, v_0, v_e)$ is a labeled tree, where
  $V$ is a set of nodes,
  $M : V \maps S$ maps nodes to labels, 
  $C : V \times \naturals \pmaps V$ maps the indexed children of nodes,
  $v_0 \in V$ is the root, and
  $v_e \in V$ is the root of expanding part of the rule.
\end{df}

If we {\em apply}, a rule $r$ on a molecule $m$, then it is extended at some node
$v \in m$. A copy of the descendants of $v_e$ will be attached to
$v$ in $m$, and the rest of the nodes in the rule have to match $v$ and above.
We call the descendants of $v_e$ as {\em expanding nodes}
and all the other nodes as {\em matching nodes}.

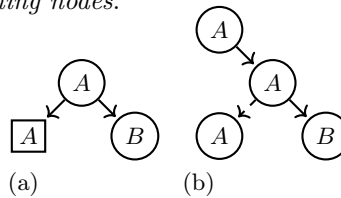
\begin{wrapfigure}{r}{0.37\textwidth}
  \vspace{-12mm}
    % \begin{minipage}{0.48\linewidth}
      \small
    % \center
  \begin{tikzpicture}[shorten >=1pt,thick,node distance=1cm,on grid]
    \node[loc] (v0) {$A$};
    \node[loc, below right of=v0] (v1) {$B$};
    \node[sqloc, below left of=v0] (v2) {$A$};
    \path[->] (v0) edge (v1);
    \path[->] (v0) edge (v2);
  \end{tikzpicture}
  % \par
  % (a)
  % \end{minipage}
  % \begin{minipage}{0.48\linewidth}
    % \small
    % \center
  \quad
  \begin{tikzpicture}[shorten >=1pt,thick,node distance=1cm,on grid]
    \node[loc] (v0) {$A$};
    \node[loc, below right of=v0] (v1) {$A$};
    \node[loc, below right of=v1] (v3) {$B$};
    \node[loc, below left of=v1] (v2) {$A$};
    \path[->] (v0) edge (v1);
    \path[->,dashed] (v1) edge (v2);
    \path[->] (v1) edge (v3);
  \end{tikzpicture}
  % \par
  \hfill (a) \hfill (b)\hfill\mbox{}
  % (b)
  % \end{minipage}
  \caption{(a) A rule. (b) An application of the rule.}
  \label{fig:exrule}
  \vspace{-6mm}
\end{wrapfigure}
\paragraph{Example:} In Figure~\ref{fig:exrule}(a), we present a rule. It has
  two kinds of nodes.
  The rule adds the square node ($v_e$).
  The circular nodes %$A$ and its right child $B$
  are the pattern, which must be present in the molecule to apply the rule.
  In Figure~\ref{fig:exrule}(b), we present an application of the rule.
  The solid tree with three nodes is the initial molecule.
  The middle node $A$ and its right child $B$ form a pattern, where the rule is applicable.
  %Upon applying the rule,
  The rule adds a left child with the label $A$ to the middle node.
  The rule is not applicable at the root $A$ due to pattern mismatch.
  % , since it has a right child $A$
  % and the combination does not match the pattern.

We naturally extend the definitions related to molecules, including $Match$ and $Copy$,
to the production rules.
Let us formally define the molecule productions using the rules.
Let $m = (V,M,C,v_0)$ be a molecule and $r = (V_r, M_r, C_r, v_{0r}, v_e)$ be a production rule.
Let $d$ be such that $v_{0r} = ancestor(r,v_e,d)$, i.e., $v_e$ is at the depth $d$ in $r$.
Let $i$ be such that $C_r(v',i) = v_{0r}$ for some $v' \in r$.
We apply $r$ on $m$ at node $v \in m$ such that $C(v,i) = \bot$.
We obtain an expanded molecule as follows.
Let $(V',M',C',v_0') = Copy(r,v_e)$.
The expanded molecule is
$
m' = (V \uplus V', M \uplus M', C \uplus C' \uplus \{(v,i) \mapsto v_0'\}, v_0)
$ if
$Match( r, v_{0r}, m', ancestor(m', v_0', d) )$ where $\uplus$ is the disjoint union.
The match condition states that after attaching the new nodes $V'$
the rule tree must be embedded in $m'$ at the $d$th ancestor of $v'_0$. 
We write $m' = Apply(m, v, r)$ to indicate the application of $r$
on molecule $m$ at node $v$ that results in $m'$.
If $r$ is not applicable at $v$, we write $Apply(m, v, r) = \bot$.
We write $m' = Apply(m, r)$ if there is a $v \in m$ such that
$m' = Apply(m,v,r)$.

Let $R$ be a set of rules.
A molecule $m$ is {\em producible} by $R$ from a set of molecules $Q$
if there is sequence of molecules $m_0,...,m_k$
such that $m_0 \in Q$, $m_k=m$, and
for each $0<i\leq k$, $m_{i} = Apply(m_{i-1},r)$ for some $r \in R$.
Let $P(Q,R)$ denote the set of molecules that are producible from
rules $R$ from a set of molecules $Q$.
We have discussed in Section~\ref{sec:bio} that all the production rules
are not applied at the same time.
The rules may live in compartments and
the rule sets of the compartments are applied one after another.
To model compartments for the rules,
let us suppose we have a sequence $R_1,...,R_k$ of set of rules.
Let $P(Q, R_1,..,R_k) = P(..P(P(Q,R_1),R_2),..,R_k)$ denoting the
trees obtained after applying the rule sets one after another.

% \subsection{Synthesis problem}
In nature, we observe a set of glycan molecules $\mu$ present in a cell.
However, we do not necessarily know the production rules to produce the molecules.
We will be developing a method to find the production rules.
The {\em synthesis problem} is to find a set $R$ of production rules
such that $\mu = P(S,R)$,
where $S$ is the set of monomers.

% %
% We can define a more general form of the synthesis problem.
% Let us suppose we are also given $k$ compartments with unknown rules.
% The goal of synthesis is to find  $R_1,...,R_k$ such that
% $\mu = P(S, R_1,..,R_k)$.
% We may relax the requirement of exactly producing $\mu$.
% We may say that molecules in $\mu$ have to be produced, but we are ok
% if subtrees of the molecules of $\mu$ are also produced.
% Formally, we may weaken the requirement to $\mu \subseteq P(Q, R_1,..,R_k) \subseteq subtree(\mu)$.

% The above are a few simplified versions of the biological problem.
% We may generalize the problem further where
% %rules are partitioned and the partitions are applied in phases one after another,
% rules are not applied exhaustively due to time constraints,
% the given $\mu$ is finite and may not be exhaustive,
% and we only know the weights of the parts of molecules in $\mu$.
% We will first present a method for solving a simplified version of the problem.
% However, our tool handles some of the above variations.
% We will discuss the variations in section~\ref{sec:variations}.

%--------------------- DO NOT ERASE BELOW THIS LINE --------------------------

%%% Local Variables: 
%%% mode: latex
%%% TeX-master: "main"
%%% End: 

% \clearpage

\section{Method for the synthesis problem}
\label{sec:algo}
In this section, we present a method to solve the synthesis problem of finding
a set of production rules that produce a given set $\mu$ of molecules.
Our synthesis method \textsc{SugarSynth} is presented in Algorithm~\ref{alg:sugar-synth}.
Here, we are considering only the single compartment case.
The generalizations are discussed in
\choosefinal{the extended version~\cite{arxiv-this}.}{ Appendix~\ref{sec:variations}.}

% At a high level, our method is iterative.
% The method first synthesizes a set of rules $R$ that can produce the molecules in input set $\mu$, i.e., $\mu \subseteq P(S,R)$.
% However, there is also an additional requirement that a molecule
% that is not in $\mu$ is not producible by the synthesized rules $R$.
% Therefore, the method looks for the producible molecules that are not in $\mu$, called
% {\em counterexample molecule}.
% If there is no such molecule, our method terminates and reports
% the synthesized rules.
% If there is a molecule in $P(S,R) - \mu$, we add constraints that says
% that look for rules
% that do not generate
% the molecule and go for another iteration.
% Now let us look at the details of the method.

\begin{algorithm}[t]
  \caption{ \textsc{SugarSynth}($\mu$, $d$, $n$)}
  \textbf{Input     :} $\mu$ : molecules, $d$ : maximum rule depth, $n$ : number of rules \\
      \textbf{Output:}  $R$: synthesized rules
  \label{alg:sugar-synth}
  \begin{algorithmic}[1]
  \State $S$ := the set of monomers appear in $\mu$, $w$ := the maximum arity of a monomer in $S$
  \State $T$ := \textsc{MakeTemplatesRule}( $S$, $d$, $w$, $n$)
  \label{line:createRtemp}
  \State tCons := \textsc{RuleTemplateCorrectness}(T)
  \label{line:ruleCorr}
  \State Let $h$ is maximum of the heights of molecules in $\mu$
  \State $\hat{m}$ := \textsc{MakeTemplateMol}( $S$, $h$, $w$ )
  \label{line:createMtemp}
  \State $mCons := \textsc{MolTemplateCorrectness}(\hat{m},\mu)$
  \label{line:molCorr}
  \State $pCons$ := $\Land_{m \in \mu}$ \textsc{EncodeProduce}(m,T)
    \label{line:molenc}
  \State $nCons := \ltrue$
  %\State $negMol$ be set of molecules := $\emptyset$
  \While{$\ltrue$} \algorithmiccomment{while True}
    \If{ $a$ = getModel( $tCons \land pCons \land nCons$ )}
    \label{line:posModel}
    \State $R$ := \textsc{readRules}($a$)
    \label{line:getR}
    \Else
       \State \textbf{ throw} Failed to synthesize the rules!
    \EndIf
    \State $rCons$ := \textsc{EncodeProduce}($\hat{m}$,Rs)
    \label{line:consNewR}
    \If{ $a$ = getModel( $mCons \land rCons$ )}
        \label{line:negModel}
        \State $m' :=$ \textsc{getNegMol}( $\hat{m}$, a)
        \State $nCons \landplus \forall \tau,cuts. \lnot \textsc{EncodeProduce}(m',T)$
        \label{line:encode-neg-mol}
        \label{line:negCons}
    \Else
       \State \Return synthesized rules $R$
    \EndIf

  \EndWhile
\end{algorithmic}
\end{algorithm}

%--------------------- DO NOT ERASE BELOW THIS LINE --------------------------

%%% Local Variables: 
%%% mode: latex
%%% TeX-master: "main"
%%% End: 

\subsection{\textsc{SugarSynth} in detail}

The method assumes that the input set $\mu$ is finite.
This is a reasonable assumption because
even if a set of rules can produce
an unbounded number of molecules,
no biology will exhibit an infinite set in a cell.
Our method
bounds the search space of production rules.
The method also takes two numbers as input:
$d$ is the maximum height of the learned rules
and 
$n$ is the maximum number of them.
% The parameters eliminate potential
% unbounded search space.
If the method fails to find production rules,
the user may call the method with larger parameters.
% However, we do not expect to find a large number of or sized rules
% because chemical reactions do not usually have many conditional
% on the precursors and do not apply large changes in a single step.
First, the method initializes $S$ with the set of monomers occurring
in $\mu$
and sets $w$ to be the maximum arity of any monomer in $S$.

We use templates to model the search space of rules.
A template is also a tree that has a depth and
the internal nodes of the tree have the same number of children.
Two variables label each node of the template.
One variable is for choosing the sugar at the node and the other is for describing
the `situation' of the node.
% The first variable at the node encodes choices of
% sugar monomers or their absence.
The domain of the first variables is $S \union \{\bot\}$.
Let $SVars$ be the unbounded set of variables with the domain.
We will use the pool of $SVars$ to add variables to the templates.

\begin{wrapfigure}{r}{0.35\textwidth}
    
        \center
  \begin{tikzpicture}[shorten >=1pt,thick,node distance=1cm,on grid,scale=0.5]
    \node[loc] (v0) {$v_0$};
    \node[sqloc, below right of=v0] (v1) {$v_e$};
    \draw [->] plot [smooth] coordinates { (v0.south)
          ($(v0.south)+(3mm,-2mm)$)
          ($(v0.south)+(10mm,1mm)$)
          % ($(v0.south)+(9mm,-6mm)$)
          % ($(v0.east)+(8mm,-8mm)$)
          (v1.north) };
    % \path[->] (v0) edge (v1);
    % \path[->,dashed] (v1) edge (v2);
    % \path[->] (v1) edge (v3);
    \fill[fill=gray,opacity=0.2] 
    (0,0) -- ++(-1.5cm,-1.7cm) -- ++(1.2cm,-1.2cm) -- ++(6cm,0cm) -- cycle;
    \fill[fill=gray,opacity=0.4] 
    (v1.center) -- ++(-1.5cm,-1.5cm) -- ++(3cm,0cm) -- cycle;
    \draw[dashed] (-1.5cm,-1.7cm) -- ++(-1.2cm,-1.2cm) -- ++(2.4cm,0cm) -- cycle;
  \end{tikzpicture}
  \caption{Parts of production rules in the rule templates.}
  \label{fig:rule-parts}
  \vspace{-8mm}
\end{wrapfigure}
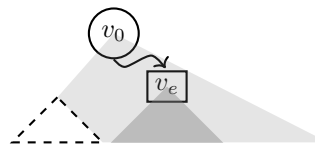

A node in a production rule can be in four situations.
In Figure~\ref{fig:rule-parts}, we illustrate the situations.
The first situation is when a node position is in the expanding part.
The dark gray nodes are the expanding nodes.
The second situation is when a node position is not in the rule.
The dashed area are the absent nodes. Among the matching nodes, we have two cases.
The third situation is when a node position is in the matching part
and has expanding descendants.
The nodes on the solid path from $v_0$ to the root $v_e$ of the expanding part are in the third situation.
% the nodes that are ancestors of expanding part among the matching nodes.
Finally, the fourth situation is the rest of the node positions
in the matching part, which is the light gray area.
% The nodes in the rest of the light gray area are in the fourth situation.
A variable is mapped to a node to encode the four situations.
Let $K =\{Expand,Absent,MatchAns,Match\}$ be the set of symbols to
indicate the set of four situations.
Let $KVars$ be the unbounded set of variables with domain $K$. 
% We will use the pool of $KVars$ to add variables to the templates.
Our templates are sufficiently expressive to cover all aspects of biology.
The templates are defined as follows.

\begin{df}
For given integers $d$ and $w$,
a {\em rule template} $t = (V, \nu, \kappa, C, v_{0r})$ is a labeled full tree with depth $d$ and 
each internal node has $w$ children, where 
$V$ is a set of nodes, $\nu : V \maps SVars$ maps nodes to distinct
sugar choice variables, 
$\kappa : V \maps KVars$ maps nodes to distinct situation variables,
$C : V \times \naturals \pmaps V$ maps the indexed children of nodes,
and
$v_{0r} \in V$ is the root of the tree.
\end{df}

For a node $v$ in a template if we assign $\kappa(v) = Absent$,
we call the node {\em absent}. Otherwise, we call the node {\em present}.
We will also be searching for the molecules that may be produced by the learned rules.
Therefore, we need to define the search space for the molecules.
We use templates for defining the search space.
We limit the template size using a parameter,
namely the height of the template.
\begin{df}
For given integers $h$ and $w$,
a {molecule template} $\hat{m} = (V, \nu, C, v_{0m})$ is a labeled full tree with height $h$ and 
each internal node has $w$ children, where 
$V$ is a set of nodes, $\nu : V \maps SVars$ maps nodes to sugar choice variables,
$C : V \times \naturals \pmaps V$ maps the indexed children of nodes,
and
$v_{0m} \in V$ is the root.
\end{df}

In the Algorithm at line~\ref{line:createRtemp}, we call \textsc{MakeTemplatesRule}( $S$, $d$, $w$, $n$) to create $n$ templates for
height $d$ and children width $w$.
% We get $d$ and $n$ as input, and
% we let $w$ be the maximum arity of any sugar that occurs in $\mu$.
Since $w$ is the maximum arity of any sugar, we can map any node to any sugar.
% We also assume that templates
% returned by $\textsc{MakeTemplatesRule}$ do not share variables.
% We obtain a production rule from a template by assigning values
% to the variables of the template.
% Not all the assignments may result in a valid rule.
Next at line~\ref{line:ruleCorr}, we will construct constraints
that encode the set of valid rules. 
A valid assignment to the variables in
a template  $t = (V, \nu, \kappa, C, r)$ must
satisfy the following six conditions.
% The first two encode the basic agreement among variables.
% The third encodes the tree structure of the rule.
% The last three encode the relative positions of the four
% parts of the rule.
\begin{enumerate}
\item If a node is present, then it is labeled with a sugar.\\
  %Otherwise, we ignore the label.
$
\Land_{s \in S} \Land_{v \in V} (\kappa(v) \neq Absent \limplies \nu(v) \neq \bot )
$
\item
  % The number of children of a node should match with the arity of the labeled sugar.
  % The following formula states that

  The children that are at greater arity than that of the label
  are absent.\\
$
\Land_{s \in S} \Land_{i \in (arity(s),w] }\Land_{v \in V} (\nu(v) = s \limplies \kappa( C(v,i) ) = Absent )
$
\item If a node is present, then the parent of the node is also present.\\
  $
  \Land_{{\text{internal node } v \in V}}
  \Land_{i\in[1,w]} (\kappa(C(v,i) ) \neq Absent \limplies \kappa( v ) \neq Absent )  
  $
\item If a node is $Expand$, then its children are also $Expand$ if present.\\
  % The following constraints state that once one reached the expanding part, there are no descendants that are matching.
  $
\Land_{i\in[1,w]} \Land_{{\text{internal node } v \in V}} (\kappa(v) = Expand \limplies \kappa(C(v,i)) \in \{Expand,Absent\}   )
$
\item If a node is $Match$, then its children are also $Match$ if present.\\
  % Similarly, the following constraints state that once one reached the matching part without expanding descendants, there are no descendants that are expanding.
$
\Land_{i\in[1,w]} \Land_{{\text{internal node } v \in V}} (\kappa(v) = Match \limplies \kappa(C(v,i)) \in \{Match,Absent\}   )
$
\item If a node is $MatchAns$, then exactly one child is $MatchAns$ or $Expand$.\\
  % The following constraints encode that there is exactly one subtree that is expanding part
  % by stating that there is no branching in $MatchAns$ nodes and they form a path.
\mbox{}\hspace{-5mm}  
$
\Land_{v \in V} (\kappa(v) = MatchAns \limplies \sum_{i\in[1,w]}( \kappa(C(v,i)) \in \{MatchAns,Expand\}) = 1
$
\end{enumerate}

The call to $\textsc{RuleTemplateCorrectness}$ at
line~\ref{line:ruleCorr} creates the above constraints
and stores them in $tCons$.
% Next, 
At line~\ref{line:createMtemp}, we use the call to \textsc{MakeTemplateMol}($S$, $h$, $w$) to
create a molecule template of height $h$ and children width $w$.
Our method searches for unwanted producible molecules up to the height of $h$,
which we set to the maximum of the heights of molecules in $\mu$.
The choice of $h$ is arbitrary.
% %
% We assume if the rules could have produced a molecule with height less than $h$,
% then it must be in $\mu$.
% On the other hand, 
% We ignore the potential formation of the large size of molecules.
% %
% Since we observe only finite molecules in any wet experiment, we may assume
% there may be large producible molecules, but we did not observe them.
% However, this point is debatable and $h$ may be viewed as
% a parameter of the whole method.
%
Similar to the rule templates, not all assignments to molecule template variables
are valid.
We add the following conditions for
valid assignments for molecule template $\hat{m} = (V, \nu, C, \hat{v_0})$.
% The first  two conditions encode the tree structure of the molecule and
% matching the number of children with label arity for each node.
% The second encodes that the height of the molecule is large enough.
% The last encodes that new molecule must not be the same as any molecule in $\mu$.
\begin{enumerate}
\item If a node is present, then the parent of the nodes is also present.\\
  $
  \Land_{i\in[1,w]}\Land_{{\text{internal node } v \in V}} (\nu(C(v,i) ) \neq \bot \limplies \nu( v ) \neq \bot )  
  $
\item The children count of a node matches with the arity of the labeled sugar.\\
  $
  \Land_{s \in S} \Land_{i \in (arity(s),w] }\Land_{v \in V} (\nu(v) = s \limplies \kappa( C(v,i) ) = Absent )
  $
% \item At least one of the leaves is present. This is for
%   avoiding the trivial solution.\\
%   $
%   \Lor_{ v \in V \text{ and } depth(v) > d' } \nu( v ) \neq \bot,
%   $
%   where $d'$ is a number close to and smaller than $height(\hat{v_0})$.
%   This allows us to look for all molecules that are close
%   to height $h$.
%   We choose $d' = h-d$, where $d$ is the height of the rule templates.
  % This $d'$ ensures that we do not miss producible molecules that
  % can not have exact height $h$ but is slightly larger than $h$.
  % Then, there must be another producible
  % molecule that is a subtree of the molecule and has height in
  % between $h$ and $h-d$
  % because the maximum depth of a rule is $d$.
\item We find a molecule that is not in $\mu$.
  We encode $\Land_{{(\_,\_,\_, v_0)\in \mu}}  Neq(\hat{v_0},v_0)$, where
   predicate $Neq$ be recursively defined as follows.\\
  $Neq(v, v') :=\; \nu(v) \neq M(v') \lor
  \Lor_{{i \in [1,w]}} Neq( C(v,i), C(v',i ) )$\\
  $Neq(v,\bot) :=\;  \nu(v) \neq \bot$
  % \begin{align*}
  % Neq(v, v') :=\;&   \nu(v) \neq M(v') \lor
  % \Lor_{\mathclap{i \in [1,w]}} Neq( C(v,i), C(v',i ) )\\
  %   Neq(v,\bot) :=\;&  \nu(v) \neq \bot
  % \end{align*}
  % The constraint is
  % % \begin{align*}
  % % \end{align*}
  % In our implementation, 
  % we have an option to also ignore the subtrees of
  % the molecules of $\mu$.
\end{enumerate}
In our method, the call to $\textsc{MolTemplateCorrectness}$ at
line~\ref{line:molCorr} generates the above constraints and stores them in
$mCons$. 

We need to encode that the rules do generate the molecules
in $\mu$ and do not generate any other.
Using procedure $\textsc{EncodeProduce}$, we generate
the corresponding constraints.
We will discuss the procedure in-depth shortly.
Let us continue with~\textsc{SugarSynth}.
At line~\ref{line:molenc}, we call $\textsc{EncodeProduce}$ for
each molecule in $\mu$ and generate constraints $pCons$ stating that
the solutions of the templates will produce the molecules in $\mu$.

After producing the constraints $tCons$, $mCons$, and $pCons$,
the method enters in the loop.
It first checks the satisfiability of conjunction
$tCons \land pCons \land nCons$, where $nCons$ is $\ltrue$ in the first iteration
and will encode constraints related to counterexample molecules.
If the conjunction is unsatisfiable, there are no rules of
the input number and height,
and the method returns failure of synthesis.
If it is satisfiable, it constructs the rules at line~\ref{line:getR} from
%the satisfying assignment $a$
and stores them in $R$.

At line~\ref{line:consNewR}, we construct constraints $rCons$ again using
$\textsc{EnocdeProduce}$ that says that template molecule $\hat{m}$
is generated by rules $R$.
We check the satisfiability of $rCons \land mCons$.
If it is not satisfiable, we have found the rules that generate exactly
the molecules in $\mu$ and the loop terminates.
Otherwise, we use the satisfying assignment to create a
counterexample molecule $m'$.
At line~\ref{line:negCons}, we add constraints to $nCons$ stating that
all possible applications of template rules in $T$ must not produce $m'$.
We use shorthand $F \landplus G$ for $ F := F \land G$.
As we will see that $\textsc{EncodeProduce}$ introduces fresh variable
maps $\tau$ and $cuts$ in the encoding.
Since we negate the returned formula by $\textsc{EncodeProduce}$ 
and then we check the satisfiability.
% the encoding will not be correct
% due to the fresh variables.
Therefore, we need to introduce universal quantifiers for the
fresh variables.
We introduce universal quantifiers over $\tau$ and $cuts$
variables before adding to $nCons$.
% Intuitively, the universal quantifiers say that all possible
% application patterns of the learned rules do not produce $m'$.
Afterwards, the loop goes to the next iteration.
Since the domain of all the variables is finite, eventually the loop must
terminate.

\subsection{\textsc{EncodeProduce} in detail}

Now let us look at the encoding generated by $\textsc{EncodeProduce}$.
The process of production of molecules adds pieces of trees
one after another.
In order to show that a molecule is producible by a set of production rules,
we need to find the nodes where the production rules are applied,
the rules that are applied on the nodes, and
the order of the application of the rules.
% The encoding is like encoding bounded executions of programs.
To model the production due to the application of the rules,
we attach three maps to the molecule nodes.
\begin{itemize}
\item Let $cuts$ map each node to a Boolean variable
indicating the node is the point where a rule is applied to
expand the molecule.
We say points of the applications of the production
rules as {\em cuts} of the tree.
% All the nodes that are in between the cuts are added in a single step.

\item Let $rmatch$ map each node to a rule indicating
  the rule that is applied to expand at the node.
  We need to match a rule to a node if it is a cut point.
  % If a node is cut point, we need to the rules that added the nodes  
\item Let $\tau$ map each node to an integer variable
  indicating the time point when the node was added to the molecule.
  Since already added nodes in a molecule decide what can be added later,
  we need to record the order of the addition.
  % If we do not capture the order in our encoding, we may wrongly declare
  % that a molecule is producible while it is not due to the matching
  % constraints.
\end{itemize}

\vspace{-3mm}
\algnewcommand{\IIf}[1]{\State\algorithmicif\ #1\ \algorithmicthen}
\algnewcommand{\EndIIf}{\unskip\ \algorithmicend\ \algorithmicif}
\setlength{\textfloatsep}{0pt}%
\begin{algorithm}[H]
  \caption{\textsc{EncodeProduce}( $m$ : molecule (template), $T$ : rule (template) )}
  \label{alg:produce}
  \begin{algorithmic}[1]
    \State \Return $
    \Land_{v\in m}
    \Land_{t \in T}
    % \Land_{{\substack{m=(V,\_,\_,\_) \\ v\in V \\ t \in T }}}
    \left[
      rmatch(v) = t \land cuts(v) \limplies \Lor_{ {\ell \in [1,d)}} \textsc{EncodeP}( v, t, \ell)
      \right]
      $
  \end{algorithmic}
  \hrule
  \vspace{1ex}
  \hrule\vspace{2pt}
  \textsc{EncodeP}( $v$, $t = (\_,\nu,\kappa,v_{0r})$, $\ell$)\hfill\mbox{}
  \vspace{2pt}\hrule
  \begin{algorithmic}[1]
    \IIf{$ancestor(v,\ell) = \bot$} \Return $\lfalse$ \EndIIf
    \State $mark := \tau(v)$,\; $c := \ltrue$, \; $v_r := v_{0r}$, \; $i = \ell$
    \While{$i > 0$}
    \label{line:encodep-while}
    \State $v' := ancestor(v,i)$
    \State $c \landplus  \kappa(v_r) = MatchAns \land \nu(v_r) = M(v') \land \tau(v') < mark$
    \label{line:encodep-ans-match}
    \For{ $j \in [1,NumberChildren(v')]$ }
    \label{line:encodep-sub-for-loop}
    \If{$C(v',j) = ancestor(v,i-1)$}
       \label{line:encodep-jth-child-cond}
       \State $v_r' := C(v_r, j)$ 
       \label{line:encodep-next-vr}
    \Else
       \State $c \landplus \textsc{MatchTree}(C(v',j), C(v_r, j),mark, \lfalse)$
       \label{line:mtree}
    \EndIf
    \EndFor
    \State $v_r := v_r'$,\; $i := i - 1$
    \label{line:encodep-update-vr}
    \EndWhile
    \label{line:encodep-end-whileloop}
    \State $c \landplus  \kappa(v_r) = Expand \land \textsc{MatchTree}(v, v_r,mark,\ltrue) \land \textsc{MatchCut}(v, v_r,\lfalse)$
    \label{line:encodep-vr-expand}
    % % \State $c \landplus   $
    % \label{line:encodep-mtree-expand}
    % % \State $c \landplus  $
    % \label{line:mcut}
    \State \Return $c$
  \end{algorithmic}
  \hrule
  \vspace{1ex}
  \hrule\vspace{2pt}
  \textsc{MatchTree}( $v$, $v_r$, $mark$, $isExpand$ )\hfill\mbox{}
  \vspace{2pt}\hrule
  \begin{algorithmic}[1]
    \IIf{$v_r = \bot$} \Return $\ltrue$ \EndIIf
    \label{line:matchtree-vr-absent}
    \IIf{$v = \bot$} \Return $\kappa(v_r) = Absent$ \EndIIf
    \label{line:matchtree-v-absent}
    \State $tCons := isExpand$ \;? $( mark \leq \tau(v) ): ( \tau(v) < mark )$
    % \If{$isExpand$}
    %   \State $tCons := ( mark \leq \tau(v) )$
    % \Else
    %   \State $tCons := ( \tau(v) < mark ) $
    % \EndIf
    \State $c := \kappa(v_r) \neq Absent \limplies tCons \land \nu(v_r)= M(v) $
    \label{line:matchtree-cons}
    % \For{ $i \in [1,NumberOfChidren(v)]$ }
    % \State\hspace{-2ex} $c \landplus \textsc{MatchTree}( C(v,i), C(v_r,i), mark, isExpand)$
    % \EndFor
    \State  $c \landplus \Land_{i \in [1,NumberOfChidren(v)]} \textsc{MatchTree}( C(v,i), C(v_r,i), mark, isExpand)$
    \label{line:matchtree-recurse}
    \State \Return $c$   
  \end{algorithmic}
  \hrule
  \vspace{1ex}
  \hrule\vspace{2pt}
  \textsc{MatchCut}( $v$, $v_r$, $ruleParentIsNotAbsent$ )\hfill\mbox{}
  \vspace{2pt}\hrule
  \begin{algorithmic}[1]
    \IIf{$v = \bot$} \Return $\ltrue$ \EndIIf
    \label{line:mcut-no-v}
    \IIf{$v_r = \bot$} \Return $parentIsNotAbsent \limplies cuts(v)$ \EndIIf
    \label{line:mcut-no-vr}
    \State $c := ruleParentIsNotAbsent \limplies ( \kappa(v_r) = Absent) = cuts(v)$
    \label{line:mcut-cons}
    % \For{ $i \in [1,NumberOfChidren(v)]$ }
    % \State \hspace{-1.5ex}$c \landplus \textsc{MatchCut}( C(v,i), C(v_r,i), \kappa(v_r) \neq Absent)$
    % \EndFor
    \State $c \landplus \Land_{i \in [1,NumberOfChidren(v)]}\textsc{MatchCut}( C(v,i), C(v_r,i), \kappa(v_r) \neq Absent)$
    \label{line:mcut-kids}
    \State \Return $cons$
  \end{algorithmic}      

\end{algorithm}

%--------------------- DO NOT ERASE BELOW THIS LINE --------------------------

%%% Local Variables:
%%% mode: latex
%%% TeX-master: "main"
%%% End:

Algorithm~\ref{alg:produce} presents function $\textsc{EncodeProduce}$
that returns the encoding.
It takes a molecule $m$ and a set of rules $T$.
Both the inputs can be template or concrete.
% The function is tolerant of the variations, but
Our presentation assumes that the molecule is concrete and
the rules are templates.
This will cover the case when the $\textsc{EncodeProduce}$ is called at
line~\ref{line:molenc} in Algorithm~\ref{alg:sugar-synth}.
However, at line \ref{line:consNewR} the molecule is a template
and the rules are concrete, which we will discuss later.
% We will discuss the handling of this situation at the end of this subsection.
$\textsc{EncodeProduce}$ uses the help of three other supporting functions,
$\textsc{EncodeP}$, $\textsc{MatchTree}$, and $\textsc{MatchCut}$.
% All of them are presented in algorithm~\ref{alg:produce}.

$\textsc{EncodeProduce}$ returns constraints stating for each node $v$
and rule template $t$,
if $v$ is at a cut and $t$ is applied at $v$, then $t$ must match
at the node $v$.
We require $rmatch(v)$ to be equal to some rule.
% Therefore, if $v$ is a cut node, some template must be applied. 
Since we are matching with a template rule, we do not know the
position of the root of expanding nodes.
We enumerate to all possible depths from $1$ to $d-1$ for finding
the root.
For each $\ell \in [1,d)$, we call $\textsc{EncodeP}( v, t, \ell)$
to construct the constraints encoding that $t$ is applied at $v$
and the depth of the root of the expanding nodes is at depth $\ell$.

In $\textsc{EncodeP}$, we can traverse up from $v$ for $\ell$ steps to
find the node to match the root of $t$.
Naturally, it returns $\lfalse$ if there
is no $\ell$th ancestor of $v$. % i.e., there is no match.
The variable $mark$ is the timestamp for node $v$.
% During matching, we will add constraints stating that the matching part of the
% template should match to the nodes added before $mark$
% and expanding part of the template should match with nodes that
% are added at or after $mark$.
The local variable $c$ collects the conjunction of constraints as they are
generated. % and $c$ is initially $\ltrue$.
The local variable $v_r$ is initially equal to the root $v_{0r}$ of $t$
and traverses the nodes in $t$.
The while loop at line~\ref{line:encodep-while} in $\textsc{EncodeP}$ starts with
the $\ell$th ancestor from $v$,
matches all the ancestors up to the parent of $v$.
As the loop traverses down, it also traverses $t$ along the matching
using variable $v_r$.
In each iteration, $v_r$ is updated to the $j$th child due to lines
\ref{line:encodep-next-vr} and \ref{line:encodep-update-vr}
if the ancestors of $v$ also traverse along with the $j$th child.
% (see if condition at line \ref{line:encodep-jth-child-cond}).

% Let us consider the body of the loop.
Let $v'$ be the $i$th ancestor at some iteration of the loop.
At line~\ref{line:encodep-ans-match}, the loop adds constraints stating that
the corresponding node $v_r$ in the template rule is
$MatchAns$, sugar matches in $v_r$ and $v'$, and
node $v'$ is added before $mark$.
The loop at line~\ref{line:encodep-sub-for-loop}, iterates over
children of $v'$.
If there is a child of $v'$ that is not along the path to
$v$, it is matched at line~\ref{line:mtree} with the corresponding
child of $v_r$ by calling $\textsc{MatchTree}$, which traverses
rule template and molecule and generate constraints to encode that their
nodes match.
We will discuss $\textsc{MatchTree}$ shortly.
% $\textsc{MatchTree}$ is also called with a third parameter
% $mark$ and a Boolean value as the fourth parameter.
% We will discuss the procedure shortly.
If the $j$th child of $v'$ is along the path to $v$,
we get node $v'_r$ for updating
$v_r$ for the next iteration of the while loop.
% After the while loop in $\textsc{EncodeP}$, $v_r$ is the
% candidate node for the root of expanding nodes.
After the while loop in $\textsc{EncodeP}$ at line \ref{line:encodep-vr-expand}, we declare $v_r$
is $Expand$,  match the node $v$
with the corresponding node $v_r$ in the template rule by calling
$\textsc{MatchTree}$, and also call to match the cut pattern at
the subtree of $v$ with the template rule.

$\textsc{MatchTree}$ is a recursive procedure and matches
sugar assignments between the molecule and the rule template.
If the rule template node $v_r$ does not exist, then
there is nothing to match and it returns $\ltrue$ at
line~\ref{line:matchtree-vr-absent}.
At line~\ref{line:matchtree-v-absent}, we encode if the molecule node $v$ does not exist, then the rule node must also
be flagged absent. %Otherwise, the match will fail.
Otherwise, we add constraints
that if $v_r$ is not absent in the template, then
the labels of $v$ and $v_r$ must match at line~\ref{line:matchtree-cons}.
At the same line, $\textsc{MatchTree}$ also adds constraints $tCons$ that
nodes are added in the molecule in the correct order.
The last two inputs of the function are timestamp $mark$ and a bit $isExpand$,
which tells us that the matching nodes should be added before or after $mark$.
The calls to $\textsc{MatchTree}$ from \textsc{EncodeP} use the
$isExpand$ appropriately.
% to indicate that nodes correspond to the
% matching part of the rule are added before and expanding part nodes
% are added later.
Afterwords at line~\ref{line:matchtree-recurse},
the procedure calls itself for the children of $v$ and $v_r$.
% to generate the same constraints for the subtree.

Once a rule is applied to a molecule, all the
nodes inside the expanding part are added together.
Therefore, there should be no cuts within the set of added nodes.
Furthermore, if there are children nodes below the added nodes,
they must be added due to the application of some other rule.
Therefore, the children are at the cut points.
We call the above requirement {\em cut pattern}. 
$\textsc{MatchCuts}$ is a recursive function over the trees, and matches
cut patterns between the molecule node $v$ and the rule template node $v_r$.
The cuts must occur whenever nodes of the rule template transition
from present to absent.
For helping to detect the transition,
the third parameter is a constraint that encodes that the parent of $v_r$
is absent or not.
% If $v$ does not exist, then there is no constraint to add therefore
% the procedure returns $\ltrue$ at line~\ref{line:mcut-no-v}.
% If $v_r$ does not exist, then it returns constraints encoding
% that if the parent of $v_r$ was not absent, then $v$ must be at a cut at line~\ref{line:mcut-no-vr}.
% If both $v$ and $v_r$ exist, then we add constraints
% that if the parent of $v_r$ is not absent, $v_r$ is
% flagged to be absent in the learned rule
% if and only if $v$ is the cut points at line~\ref{line:mcut-cons}.
% $\textsc{MatchCuts}$ recursively traverses the children of $v$ and $v_r$
% and generates the constraints for the cut pattern.
% The constraints added by the procedure are guarded by the presence of parents
% because we need to enable constraints only when there is a possibility of transition
% from presence to absence.
The call to $\textsc{MatchCuts}$ in \textsc{EncodeP} passes $\lfalse$ as the third input.
Even if the parent of $v_r$ is present,
$v_r$ is a cut point adding cut pattern constraints for the node will create inconsistency.
Therefore, we are passing $\lfalse$.

In the case when we call \textsc{EncodeProduce} with template molecule and
concrete rules.
We need to swap the roles of $\nu$ and $M$ at their occurrences
at line \ref{line:encodep-ans-match} in \textsc{EncodeP} and
line \ref{line:matchtree-cons} in \textsc{MatchTree}.
Furthermore, $\kappa(v_r)$ is a concrete value depending on the situation of $v_r$ in
the rule.
In the functions, the variable is to be replaced by the
evaluated value of $\kappa(v_r)$.

\begin{theorem}
  If \textsc{SugarSynth}($\mu$, $d$, $n$) returns a set of rules $R$, then if $R$ produces a molecule that has depth less than $h+1$, the it is in $\mu$ (soundness).
  If it fails to synthesize rules, there are no rule set from the budget of depth $d$ and number $n$ (completeness).
\end{theorem}

The soundness is true due to the construction of constraints.
Our algorithm always terminates. At each iteration, at least one solution, i.e,
a set of rules is discarded.
In fact, we discard many because in each iteration we reject production of a counterexample molecule, but this may discard whole range of rule sets. 
Since the set of all possible rules is finite, we
guarantee termination.
Therefore, the completeness holds.
In the worst case, the entire search space is explored.

% On the other hand,
% if $v_r$ is also present, then $v$ must not be any cut point.

%--------------------- DO NOT ERASE BELOW THIS LINE --------------------------

%%% Local Variables: 
%%% mode: latex
%%% TeX-master: "main"
%%% End: 

\section{Experiments}
\label{sec:experiments}
In this section, we present our implementation of the proposed method.
% %
% We also apply our tool on several data sets and illustrate the usability of our method.

{\em Implementation:} 
We have implemented method \textsc{SugarSynth} in a prototype tool {\ourtool}.
The tool, written in {\tt C++}, uses {\zthree}\cite{z3} as the SMT solver
to discharge the generated satisfiability queries. All the experiments have been conducted on a laptop with 8GB (1x8GB) DDR4 at 2400MHz.
%
% We have developed a specialized file format for giving input to~\ourtool.
% %
% The format allows one to declare monomers, list molecules over the monomers
% in the SMTLIB like format,
% and pass the input parameters of the method.

{\em Benchmark: }
We have applied our tool to three sets of real data (D1, D2, D3, D4) and two sets of synthetic data (D5, D6). The molecules have been obtained from
respiratory mucins of a cystic fibrosis patient (D1),
horse chorionic gonadotropin (D2), SARS-CoV-2 spike protein T323/S325 (D3), and human chorionic gonadotropin from a cancer cell line (D4)~\cite{Jaiman2018,10.1093/glycob/cwaa042}.
The availability of clean data, where we are clear about the source
% and the collection
% methodology used
, limits our choices.

% horse chorionic gonadotropin (D2),
% SARS-COV-2 (D3),
% 
% and
% synthetic data (D5)~\cite{}.

\begin{table}[t]
  \centering
  \tiny
  \begin{minipage}{0.48\linewidth}
  \begin{tabular}[t]{|c|c|c|c|c|c|c|}\hline
     & \#mol- & \#Rules & Rule  & \#Comp- & success? & Time \\
         & ecules   &         & depth & artments &          & (in secs.) \\\hline
    %%cf-mucin.sugar%%
         &   & 7  & 3 & 1 & Yes &  3.02 \\\cline{3-7}
    D1   & 6 & 7  & 4 & 2 & Yes & {\bf 1.60}  \\\cline{3-7}
         &   & 6  & 3 & 3 & Yes & 9.36  \\\hline
         
    %%horse-cg.sugar%%
         &   & 7  & 3 & 2 & Yes & 14.37  \\\cline{3-7}
    D2   & 3 & 5  & 3 & 2 & Yes & {\bf 7.97}  \\\cline{3-7}
         &   & 5  & 3 & 3 & Yes  &  13.42 \\\hline
         
    %%sars-cov2.sugar%%     
         &   & 6  & 4  & 2  & Yes & 1.02  \\\cline{3-7}
    D3   & 6 & 5  & 2 & 1 & Yes & {\bf 0.57}  \\\cline{3-7}
         &   & 5  & 4 & 1 & Yes  &  0.71 \\\hline
         
    %%human-cg.sugar%%     
         &   & 8  & 4  & 1  & Yes & 4.35  \\\cline{3-7}
    D4   & 3 & 6  & 3 & 1 & Yes & {\bf 0.85}  \\\cline{3-7}
         &   & 6  & 2 & 2 & No  &  1.17 \\\hline
         
    % %%D4.sugar%     
    %      &   & 6  & 2 & 1 & No &  0.64 \\\cline{3-7}
    % D5   & 3 & 7  & 2 & 1 & Yes & {\bf 0.72}  \\\cline{3-7}
    %      &   & 8  & 4 & 1 & Yes  &  2.39 \\\hline
         
    % %%D5.sugar%     
    %      &   & 5  & 3 & 2 & Yes &  0.86 \\\cline{3-7}
    % D6   & 2 & 5  & 3 & 1 & Yes & {\bf 0.73}  \\\cline{3-7}
    %      &   & 4  & 2 & 1 & No  & 0.69  \\\hline
  \end{tabular}    
  \end{minipage}
  \begin{minipage}{0.48\linewidth}

  \begin{tabular}[t]{|c|c|c|c|c|c|c|}\hline
     & \#mol- & \#Rules & Rule  & \#Comp- & success? & Time \\
         & ecules   &         & depth & artments &          & (in secs.) \\\hline
    % %%cf-mucin.sugar%%
    %      &   & 7  & 3 & 1 & Yes &  3.02 \\\cline{3-7}
    % D1   & 6 & 7  & 4 & 2 & Yes & {\bf 1.60}  \\\cline{3-7}
    %      &   & 6  & 3 & 3 & Yes & 9.36  \\\hline
         
    % %%horse-cg.sugar%%
    %      &   & 7  & 3 & 2 & Yes & 14.37  \\\cline{3-7}
    % D2   & 3 & 5  & 3 & 2 & Yes & {\bf 7.97}  \\\cline{3-7}
    %      &   & 5  & 3 & 3 & Yes  &  13.42 \\\hline
         
    % %%sars-cov2.sugar%%     
    %      &   & 6  & 4  & 2  & Yes & 1.02  \\\cline{3-7}
    % D3   & 6 & 5  & 2 & 1 & Yes & {\bf 0.57}  \\\cline{3-7}
    %      &   & 5  & 4 & 1 & Yes  &  0.71 \\\hline
         
    % %%human-cg.sugar%%     
    %      &   & 8  & 4  & 1  & Yes & 4.35  \\\cline{3-7}
    % D4   & 3 & 6  & 3 & 1 & Yes & {\bf 0.85}  \\\cline{3-7}
    %      &   & 6  & 2 & 2 & No  &  1.17 \\\hline
         
    %%D4.sugar%     
         &   & 6  & 2 & 1 & No &  0.64 \\\cline{3-7}
    D5   & 3 & 7  & 2 & 1 & Yes & {\bf 0.72}  \\\cline{3-7}
         &   & 8  & 4 & 1 & Yes  &  2.39 \\\hline
         
    %%D5.sugar%     
         &   & 5  & 3 & 2 & Yes &  0.86 \\\cline{3-7}
    D6   & 2 & 5  & 3 & 1 & Yes & {\bf 0.73}  \\\cline{3-7}
         &   & 4  & 2 & 1 & No  & 0.69  \\\hline
         
    %%D7.sugar%     
         &   & 5  & 3 & 2 & Yes &  0.72 \\\cline{3-7}
    D7   & 3 & 5  & 3 & 1 & Yes & {\bf 0.65}  \\\cline{3-7}
         &   & 6  & 2 & 1 & No  & 0.69  \\\hline
         
    %%hydra.sugar%     
         &   & 4  & 3 & 2 & No &  0.79 \\\cline{3-7}
    D8   & 3 & 5  & 3 & 2 & Yes & {\bf 0.84}  \\\cline{3-7}
         &   & 8  & 4 & 3 & Yes  & 1.53   \\\hline
  \end{tabular}
    
  \end{minipage}
  \caption{Results of applying \ourtool on data sets.
    %Bold faced numbers are the most optimal timings.
  }
  \label{tab:results}
  % \vspace{-9mm}
\end{table}
%%% Local Variables:
%%% mode: latex
%%% TeX-master: "main"
%%% End:

{\em Results:}
We have applied \ourtool~on the data set. For each data set, we choose several
parameter combinations to illustrate the relative performance of the tool.
If we did not budget large enough parameters such as the size of unknown molecules, number of rules etc, then the tool fails to
synthesize the rules.
We present the rules learned after giving minimum resources. However, the set of rules reported need not be either unique or minimal.
% Our tool gave output that matched with the analysis of~\cite{Jaiman2018}
% in reasonable time.

For D1, we synthesize the rules in 1.60 seconds. Even by reducing the first two parameters, we were able to synthesize the rules but it took longer time.
Giving an extra compartment in the third row did not impact the performance.  We also observe the trade-off between the number of compartments
and the depth of the rules at the second and third row of D1 and how it impacts performance.
We synthesize the rules for D2, involving runaway reactions,  in 7.97 seconds.
For D2, we also learned large rules, i.e., they are adding many nodes at a single time.
We synthesize the rules for D3 in 0.57 seconds. We synthesize the rules for D4, which is also our motivating example, in 0.85 seconds.
However, if we use too many resources or too little, the tool runs for a long time as the search in combinatorial space is highly sensitive to the parameters.
The synthesis for D5 takes 0.72 seconds. We can observe that by reducing the number of rules to learn, the tool fails to learn rules as it required minimum 7 rules.
Our synthesized production is in line with the reported rules
in the literature~\cite{Jaiman2018}.
\choosefinal{}{
More experiments on the variations of the problem
are included in the Appendix~\ref{sec:ex-variants}.} %\hyperref[sec:appendix]{}.

% The results suggest that the tool is potentially applicable to larger data sets.

% Currently, the tool does not have any objective function to optimize the synthesized rules.
% We are hoping to see more data sets such that we can develop a clearer picture
% for a reasonable objective function.

% ==============================================================================

% Our synthesized production rules for the first three data sets
% match with the reported rules in the literature.
% For the fourth, there is no hypothesis available to us.

% with
% the variants discussed earlier.
% However, we have not fully modeled all the information
% and biological intuition --- for example, the objective function for the rules.

% As we have discussed earlier, we have implemented variants that encode 

%--------------------- DO NOT ERASE BELOW THIS LINE --------------------------

%%% Local Variables:
%%% mode: latex
%%% TeX-master: "main"
%%% End:

% \section{Related work}
% \label{sec:related}
% \input{related.tex}

\section{Conclusion and future work}
% ##limitations in the model because of limited information,
% a)the boundary between two compartments is fuzzy, 
% b) the growths of two branches are correlated/ not correlated..and some more points like these, for which wealth of data is not available, hence the model was unable to incorporate the specifics
\label{sec:conclusion}
% contributions
% 
% In this paper,
We have presented a novel method for synthesizing production rules for glycans.
% \textsc{SugarSynth} is a CEGIS like iterative method.
We have applied our method to real-world data sets. % to illustrate that our method
% can find the rules in scenarios.
The viability of the synthesized rules can only be verified by conducting wet experiments.
We are planning to work in biological labs to check the viability of the solutions
found by our method.
Our method is the {\em first} to apply formal methods for the synthesis problem.
We have opened a new direction for the application of formal methods in biology.

\nocite{Charles2013}
\bibliographystyle{unsrt}
\bibliography{biblio}

\choosefinal{}{
\appendix

% \section{More details on Biology}

\section{Extended related work}
\label{sec:related}

% Success of formal methods
Formal methods have successful in applications to a vast range of
problems from the analysis of systems.
In particular,
the verification of hardware designs
using SAT solver based model-checking has been applied
to industrial-scale problems~\cite{biere1999symbolic2}.
The verification of software systems is a much harder problem, and
many methods like bounded model-checking~\cite{biere2003bounded},
abstract interpretation~\cite{lattice77}, and
counterexample guided abstraction refinement (CEGAR)~\cite{ClarkeCEGAR} have
been successful for the problem.

Synthesis of programs that implements a given behavior
has been a focus of research for a while~\cite{PnueliSynthesis}.
In recent years, the synthesis of programs using SAT/SMT solvers
has gained momentum.
The approach encodes the search of a program that exhibits a certain
behavior into a satisfiability problem.
The solvers attempt to find a solution to the satisfiability problem.
The solution is the synthesized program.
In~\cite{SrivastavaSynthesis,Solar-Lezama2005},
the method has been successfully applied
to find programs that satisfy quantified specifications like sorting
programs, and have missing ``holes'' and need proper implementations
for them.
In~\cite{exampleSynth},
a set of pairs of input and output examples is the specification of a program,
and the synthesis method searches for programs in a space defined by a
template~\cite{sygus}.

% Boolean networks

% Program synthesis
Since the formal methods have been exhibiting effectiveness in the vast
range of problems, the
systems biology community has been applying the methods for various
biological problems~\cite{fisher2007executable}.
The formal methods approach is distinctly different from the statistical
approach traditionally used in biology.
In formal methods, we aim to match precisely the expected behavior rather
than learn approximate artifacts from biological data.
The use of formal methods belongs to two broad categories:
the analysis of biological models and the synthesis of the models.

The key focus has been to model gene regulatory networks (GRNs), since
they are the core of the central dogma of biology.
Boolean networks are often used to model
GRNs to find stable points or attractors, i.e., nodes in the networks
where a system eventually reaches no matter where it starts~\cite{wang2012BooleanOverview}.
% In~\cite{boolean-networks-stabilty}, the networks are used to evaluate
% stability of GRNs under mutations.
The Boolean networks are not often sufficient to adequately model the behavior
of GRNs.
The continuous-time Markov chains (CTMCs) are used to bring in the aspects of
timing and probabilistic constraints to GRNs.
The transient behaviors of GRNs are also crucial for various aspects
of biology.
% For example, the decision making in some cells is based on
% response timing to stimulus~\cite{response-timing-paper}.
In~\cite{delayedCTMC}, a method based on
model-checking is used to estimate the time evolution of the probability
distributions over states of GRNs.

There are many biological processes where we can observe the system behavior,
but the exact mechanisms of the processes are not known.
Recently, the methods for formal synthesis are finding their applications in biology
\cite{dunn2014defining,xuPluripotency,booleanModelKarp13,paoletti2014analyzing}.
In~\cite{koksal2013synthesis},
GRNs with unknown interactions are modeled using
an automata-like model with missing components
and the specification is the variations of the behavior of the system under mutations.
Their approach uses a counterexample-guided inductive synthesis (CEGIS) based algorithm~\cite{cegis}.
CEGIS is a framework for synthesis. It first finds a program that may satisfy `samples' from
the specification. If the synthesized program satisfies the specifications, the CEGIS terminates.
Otherwise, the method learns a new sample where the program violates the specifications.
It adds the sample to the set and goes to the next iteration.
Typically, the constraint solvers find the programs during intermediate steps of CEGIS by
solving the generated queries.
Our method follows a similar pattern, where
a set of sampled observations is a specification, i.e., output molecules,
and we need to find the governing programs, i.e., production rules.
In a subsequent work~\cite{fisher2015synthesising}, a Boolean network model is used,
the functions attached to the nodes are considered unknown, and
a different kind of data sets provides the behavior specification.
In this approach, their earlier method is adopted to work in the new situation.

As far as we know, there has been no similar computation based analysis of glycan production rules.
However, there has been a theoretical analysis of the properties of the production
rules in~\cite{Jaiman2018}. The work identifies the conditions for the production
of a finite set of molecules and the cases of unique production methods.
In our work, we are taking the computational approach of synthesis from formal methods
instead of looking for the theoretical conditions.

\section{Variations of the synthesis problem}
\label{sec:variations}
We have presented a simplified version of
the biological problem. However, in real settings, we often encounter many more variants which require us to support additional constraints to model the set of molecules and possible rules in that particular problem. Hence, we developed the following variations of the method described in the main text to support more realistic problems.
% However, the biological problem has more details.
% We have developed variations of the above method to support more realistic problems.

% \subsection{Compartments}

{\em Compartments : } The production rules may live in different compartments.
% The compartments are ordered.
A molecule moves from one compartment to the next.
% The rules of the current compartment apply to the molecule.
% We have presented the formal description of the compartments in
% section~\ref{sec:model}.
To support the compartments, we take an additional integer
input $k$ in $\textsc{SugarSynth}$ to indicate the maximum number
of compartments.
We construct each template rule with a new variable with domain $[1,k]$.
Let $v_r$ be a node of some template rule. We write $compart(v_r)$
for the compartment variable for the template.

We will alter $\textsc{EncodeProduce}$ and its sub-procedures to
ensure that they enforce the compartment order.
We need to encode that a rule is applied when all the
pattern nodes were added in the current or earlier compartment.
% For the encoding, we add a new map $compart$ that maps
% nodes to variables with domain $[1,k]$.
We modify the function $\textsc{MatchTree}$ by replacing
$tCons$ assignment by the following code.\\
\begin{minipage}{1.0\linewidth}
\begin{algorithmic}[1]
  \setcounterref{ALG@line}{line:matchtree-v-absent}
  \State $tCons := isExpand \;?\; ( mark \leq \tau(v)  \land compart(v_r) = compart(v) )$\par
  \mbox{}\qquad\qquad\hspace{10mm} $:( \tau(v) < mark  \land compart(v_r) \geq compart(v) )$
\end{algorithmic}
\end{minipage}
Similarly, we modify the function
$\textsc{EncodeP}$ by inserting the following line after~\ref{line:encodep-ans-match}.\\
\begin{minipage}{1.0\linewidth}
\begin{algorithmic}[1]
  \setcounterref{ALG@line}{line:encodep-ans-match}
  \State $c \landplus compart(v_r) \geq compart(v')$
\end{algorithmic}  
\end{minipage}

\paragraph{Fast and slow reactions:}
There is a rate associated with chemical reactions. We abstract this by defining slow and fast rules.
The fast rules dominate the slow rule.
A slow rule can occur only when no other fast rule is able to extend the molecule in that compartment.
Let us define function $\textsc{EncodeP}^-$, which generates constraints as $\textsc{EncodeP}$ expect
line~\ref{line:encodep-vr-expand} is missing, i.e., we do not analyze expand part.
% During the stay of a molecule in a compartment, it could be observed that some reactions were dominating than others. This can be explained by characterizing reactions as either slow or fast.
% We can now define \textsc{Extend} recursively as follows:
% $  \textsc{Extends}(r, m) :=\;   Apply(m,r) \lor \Lor_{{i \in [1,w]}} \textsc{Extends}( r, C(m,i ) )$\\
% $\textsc{Extends}(r, \bot) :=\;  \nu(v) \neq \bot$\\
We now modify $\textsc{EncodeP}$ after~\ref{line:encodep-vr-expand} to support fast reactions:\\
\begin{minipage}{1.0\linewidth}
\begin{algorithmic}[1]
  % \vspace{1ex}
  \setcounterref{ALG@line}{line:encodep-vr-expand}
  \State $c \landplus \lnot \textsc{Fast}(v_r) \implies 
  \Land_{t \in T} (\textsc{Fast}(t) \implies \lnot\lor_{ {\ell \in [1,d)}} \textsc{EncodeP}^-( v, t, \ell))$
  % \vspace{1ex}
\end{algorithmic}
  % \vspace{-1mm}
\end{minipage}
Here, \textsc{Fast} sets the constraint on the rule to be fast.
A negative molecule which is a proper subtree of an input molecule will cease to be negative if there is any fast reaction that is able to extend it as fast reactions happen aggressively and can make partial molecules complete. Due to limited space, we will not present the exact constraints.

% Therefore, we modify the function $\textsc{SugarSynth}$ after~\ref{line:encode-neg-mol}:
% \begin{algorithmic}[1]
%   \setcounterref{ALG@line}{line:encode-neg-mol}
%   \State $nCons \landplus
%   \lnot  \Lor_{r \in R} \textsc{Fast}(r) \land \textsc{Extends}(r,m')$
% \end{algorithmic}
% Here, \textsc{Fast} constraints the rule to be fast. The constraints state that none of the rules should be both fast and able to extend the negative molecule received from the model.

{\em Unbounded molecules: }
We only observe a finite set of molecules in cells.
However, a set of production rules may be capable of producing an unboundedly large number of molecules.
In such cases, rules produce molecules that have repeating patterns of a subtree while rest of the tree being exactly same as one of the input molecules. The rules may be acceptable in some biological settings. We modify constraints to not declare such molecules as negative.
Let us suppose we have a repeat pattern of depth $d$ with $r$ repetitions.
Let $v$ be the node in template molecule where repetition has begin.
We define $\textsc{RepeatHeads}(v,r,d)$ that returns nodes $v_0,.... v_r$ such that
$v_1 = v$, $v_i$ is the $d$th ancestor of $v_{i+1}$.
% Those rules should in the same compartment and are slow.
Let us define \textsc{Repeat} and \textsc{Exact}, which encodes that the trees rooted at $v_{i}$s repeat.\\
$\textsc{Repeat}([v_0,...,v_r], v') :=  \textsc{Exact}(v',v_r,\bot) \land \Land_{i\in[0,r-1]} \textsc{Exact}( v_i, v_{i+1},v_{i+1})$\\
% $\textsc{Exact}(v,v') := (M(v) = M(v')) \land \Land_{i \in [1,w]} \textsc{Exact}(C(v,i),C(v',i))$ \\
% $\textsc{Exact}(\bot,v') := \bot, \textsc{Exact}(v,\bot) := \bot, \textsc{Exact}(\bot,\bot) := \bot$\\
$\textsc{Exact}(\bot,v',\_) := \lfalse, \textsc{Exact}(v,\bot,\_) := \lfalse,$\\
$ \textsc{Exact}(\bot,\bot,\_) := \ltrue,\textsc{Exact}(v_s,\_,v_s) := \ltrue$\\
$\textsc{Exact}(v,v',v_s) := (M(v) = M(v')) \land \Land_{i \in [1,w]} \textsc{Exact}(C(v,i),C(v',i),v_s)$\\
We modify constraints of $ \textsc{MolTemplateCorrectness}(\hat{m},\mu)$, which encodes that negative molecules are not in $\mu$. We replace the definition of $Neq$ as follows.\\
$Neq(v, v') :=\; (\nu(v) \neq M(v') \land \lor_{r\in[1,r_0],d\in[1,d_0]}\textsc{Repeat}(\textsc{RepeatHeads}(v,r,d), v')) $\\
\mbox{}\hspace{30mm}$\lor \Lor_{{i \in [1,w]}} Neq( C(v,i), C(v',i ) ),$\\
where $d_0$ and $r_0$ are limits on the depth of the repeating subtrees and the number of repetitions respectively.
The change will accept molecules with repeating patterns as positive samples.

\paragraph{Non-monotonic rules :}
Some production rules can not be applied if another node is present in a sibling.
We call such rules non-monotonic because it may get disabled as the molecule grows.
This feature of rules helps in producing an exact set of desired molecules.
We add an extra bit on each node of rule
template called $\textsc{HardEnds}$.
If the node is absent, its parent is present, and  $\textsc{HardEnds}$ bit is true, then
no node must be present in the matching pattern at the time of the application of the rule.
We modify the function $\textsc{MatchTree}$ by inserting the following constraints after~\ref{line:matchtree-cons}:
\begin{minipage}{1.0\linewidth}
\begin{algorithmic}[1]
  \vspace{1ex}
  \setcounterref{ALG@line}{line:matchtree-cons}
  \State $c \landplus \textsc{HardEnds}(v_r) \implies (mark \leq \tau(v))$
\end{algorithmic}  
\end{minipage}
The constraints state that if the applicable rule has \textsc{HardEnds}, then it has to be added at a time earlier than the current time of the molecule, effectively restricting the addition of further rules.

%--------------------- DO NOT ERASE BELOW THIS LINE --------------------------

%%% Local Variables:
%%% mode: latex
%%% TeX-master: "main"
%%% End:

\section{Experiments with variations of the synthesis problem}
\label{sec:ex-variants}
We conducted separate experiments for the variations of the synthesis problem - non-monotonic rules and unbounded molecules using synthetic data.
\begin{table}
  \centering
  \begin{tabular}{|c|c|c|c|c|c|c|}\hline
     & \#molecules& \#Rules & Rule depth & \#Compartments & success? & Time (in secs.) 
          \\\hline
    %%cf-mucin.sugar%%
         &   & 6  & 2 & 1 & Yes &  2.85 \\\cline{3-7}
    N1   & 6 & 6  & 2 & 5 & Yes & {\bf 1.07}  \\\cline{3-7}
         &   & 5  & 3 & 1 & No & 0.81  \\\hline
         
    %%horse-cg.sugar%%
         &   & 6  & 2 & 1 & Yes & 1.02  \\\cline{3-7}
    N2   & 6 & 6  & 2 & 5 & Yes & {\bf 0.76}  \\\cline{3-7}
         &   & 6  & 4 & 1 & Yes & 0.82 \\\hline
         
    %%sars-cov2.sugar%%     
         &   & 9  & 2  & 1  & Yes & 3.18  \\\cline{3-7}
    N3   & 9 & 8  & 2 & 1 & Yes & {\bf 2.39}  \\\cline{3-7}
         &   & 8  & 3 & 2 & Yes  &  16.69 \\\hline
         
    %%human-cg.sugar%%     
         &   & 6  & 3  & 1  & Yes & 73.61  \\\cline{3-7}
    R1   & 5 & 5  & 3 & 2 & Yes & {\bf 47.72}  \\\cline{3-7}
         &   & 4  & 4 & 1 & No  &  16.99 \\\hline
         
    % %%D4.sugar%     
         &   & 6  & 3 & 3 & Yes &  5.25 \\\cline{3-7}
    R2   & 8 & 4  & 3 & 2 & Yes & {\bf 3.44}  \\\cline{3-7}
         &   & 4  & 3 & 5 & Yes  &  4.70 \\\hline
         
    % %%D5.sugar%     
         &   & 7  & 3 & 2 & Yes &  6.85 \\\cline{3-7}
    R3   & 8 & 7  & 3 & 4 & Yes & {\bf 2.09}  \\\cline{3-7}
         &   & 6  & 3 & 1 & No  & 0.77  \\\hline
  \end{tabular}   
  \end{table}
 We applied our tool on the synthetic datasets N1, N2 and N3 which required some rules to have $\textsc{HardEnds}$. There is a clear trend of increasing time as the size of our dataset increases. Between N1 and N2 which have the same number of molecules in the dataset, N1 has bulkier molecules and hence the time taken is higher. Reducing the number of compartments increases the time slightly however is successful in both N1 and N2. By Reducing the number of rules to be learnt, the tool fails in N1 as it required a minimum of 6 rules. Rules produced in these datasets have $\textsc{HardEnds}$ in the appropriate places to ensure that the exact set of molecules can be made.

  To demonstrate the working of our tool in case of runaway reactions or datasets with unbounded molecules, we applied our tool on the synthetic datasets R1, R2 and R3. The molecules in these datasets have repeating patterns and we observe the synthesized rules to capture this pattern. The constraints added as part of unbounded molecules take care not to declare the molecules created by the repeated application of these rules to be negative. The time taken for synthesizing the rules is particularly high for R1 as the molecules in the dataset are relatively bulky. Increasing the number of compartments or Increasing the number of rules to learn increases the time in R2. The search in the combinatorial space is expensive, hence the time taken is more. 
  
\section{Implementation optimizations}

% \subsection{Optimizations}
We have incorporated several optimizations to gain efficiency or reduce
the number of iterations in the method. Let us discuss some of the remarkable
optimizations.

\paragraph{Ordered templates}
Our method constructs a list of templates.
Since the templates are symmetric, there is a potential for unnecessary
iterations.
Let us suppose the method rejects a set of learned rules.
In the subsequent queries include the constraints that reject the set of rules,
the solvers may choose the same set of rules again by permuting the mapping
from the templates to the learned rule.
To avoid these iterations, we define an arbitrary total order over the rules.
We add ordering constraints stating that any assignments to the list of templates should produce
ordered rules with respect to the total order.

\paragraph{Quantifier instantiations}
The constraints for rejecting the counterexample molecule is universally quantified at line~\ref{line:negCons} in Algorithm~\ref{alg:sugar-synth}.
For any assignment, the solver needs to try sufficiently many instantiations of the quantifiers
to ensure that the query is not unsatisfiable.
The instantiations sufficiently slow down the solving process.
We assist the solver by also adding an instantiation of the quantifiers that is equal
to the values of the corresponding variables in the assignment $a$ at line~\ref{line:negModel}.
We observe that for some inputs only adding the instantiations and not the universally quantified
formula is more effective.

\paragraph{Constraints for counterexample molecules}
In the presentation of \textsc{SugarSynth}, we construct constraints of counterexample molecules
at line~\ref{line:consNewR} in each iteration.
The repeated work is impractical because the formula management system of Z3 will be overwhelmed by
the construction of many terms repeatedly. In our implementation, we construct the constraints
once outside the while loop.
We pass the templates as the second parameter instead of concrete rules to~\textsc{EncodeProduce}.
Later at the solving time in line~\ref{line:negModel}, constraints are added to assign values
of the template variables that were returned by the solver at line~\ref{line:posModel}.
Due to the assignments, the templates become concrete rules in the context of solving.

\section{Features to support in future}

\subsection{Full organism data vs single cell data}

The experiments that observe the set of glycan molecules are of two kinds. In one kind, we isolate a single cell type organism and identify all the glycans present in the cells. In our presentation, we have assumed this kind of source of data. However, the experiments are difficult to conduct.
In another and more convenient way, we smash the whole organism and identify all the glycans present in all the cells.
So the information that which glycans are coming from which
cells is unknown.
In this situation, the synthesis has another task to map the molecules
in $\mu$ to several cell types.
A glycan may be present in multiple cell types.

Our method is easily adapted to consider this kind of data.
We search for a small covering set of subsets of
$\powerset{\mu}$ such that each set in the cover allows a small
set of production rules.
The cover sets indicate the different types of cells in terms of
the presence of glycans.
This variation makes the problem particularly hard.
There can be a potentially large number of covering subsets.
So far, we have not encountered a large enough data set such that
we apply the variation effectively.

\subsection{Incomplete data}

The experiments are imperfect. They may not detect all possible glycans
in a cell.
We may need to leave the possibility of allowing a few more molecules to be
producible beyond $\mu$.
We replace the no extra molecule constraint by a constraint that allows molecules
that are `similar' to molecules in $\mu$.
However, we could not imagine any measure of similarity that naturally stems from
biological intuition.
For now, our tool supports the count of the number of monomer differences
as a measure of similarity.

%--------------------- DO NOT ERASE BELOW THIS LINE --------------------------

%%% Local Variables:
%%% mode: latex
%%% TeX-master: "main"
%%% End:

}

% 
% To resolve this paradox we borrow from the field of
% al-gorithmic self-assembly, which explores how small buildingblocks
% with stochastic local interactions assemble into various global
% configurations [23]–[25].  A central inverse problemin self-assembly
% is to design building blocks that assembleinto a target shape.  This
% process has been studied using thetheoretical framework of Wang tiles:
% square building blockswith colored sides that stick to one another
% along sides withmatching colors [26].  For efficient assembly of Wang
% tiles,their coloring must be cleverly chosen so the target shape isthe
% unique terminal output of the stochastic assembly process[27].   That
% is, their assembly must be algorithmic.  Glycans may be considered a
% natural realization of the Wang construct,with monomers acting like
% tiles whose stickiness is encodedby GTase enzymes.  If glycan
% biosynthesis could be madealgorithmic, cells could suppress unwanted
% byproducts andgenerate only desired glycan oligomers with high yield.

\end{document}